# Improved sterile neutrino constraints from the STEREO experiment with 179 days of reactor-on data

H. Almazán,[1] L. Bernard,[2,‡] A. Blanchet,[3,§] A. Bonhomme,[1,3] C. Buck,[1] P. del Amo Sanchez,[4] I. El Atmani,[3,‖] J. Haser,[1] F. Kandzia,[5] S. Kox,[2] L. Labit,[4] J. Lamblin,[2] A. Letourneau,[3] D. Lhuillier,[3] M. Licciardi,[2] M. Lindner,[1] T. Materna,[3] A. Minotti,[3,¶] H. Pessard,[4] J.-S. Réal,[2] C. Roca,[1] R. Rogly,[3] T. Salagnac,[2,**] V. Savu,[3] S. Schoppmann,[1,*] V. Sergeyeva,[4,††] T. Soldner,[5] A. Stutz,[2] and M. Vialat[5]

(STEREO Collaboration)[†]

[1]*Max-Planck-Institut für Kernphysik, Saupfercheckweg 1, 69117 Heidelberg, Germany*
[2]*Université Grenoble Alpes, CNRS, Grenoble INP, LPSC-IN2P3, 38000 Grenoble, France*
[3]*IRFU, CEA, Université Paris-Saclay, 91191 Gif-sur-Yvette, France*
[4]*Université Grenoble Alpes, Université Savoie Mont Blanc, CNRS/IN2P3, LAPP, 74000 Annecy, France*
[5]*Institut Laue-Langevin, CS 20156, 38042 Grenoble Cedex 9, France*

The STEREO experiment is a very short baseline reactor antineutrino experiment. It is designed to test the hypothesis of light sterile neutrinos being the cause of a deficit of the observed antineutrino interaction rate at short baselines with respect to the predicted rate, known as the reactor antineutrino anomaly. The STEREO experiment measures the antineutrino energy spectrum in six identical detector cells covering baselines between 9 and 11 m from the compact core of the ILL research reactor. In this article, results from 179 days of reactor turned on and 235 days of reactor turned off are reported at a high degree of detail. The current results include improvements in the modelling of detector optical properties and the γ-cascade after neutron captures by gadolinium, the treatment of backgrounds, and the statistical method of the oscillation analysis. Using a direct comparison between antineutrino spectra of all cells, largely independent of any flux prediction, we find the data compatible with the null oscillation hypothesis. The best-fit point of the reactor antineutrino anomaly is rejected at more than 99.9% C.L.

*stefan.schoppmann@mpi-hd.mpg.de
†http://www.stereo-experiment.org.
‡Present Address: Ecole Polytechnique, CNRS/IN2P3, Laboratoire Leprince-Ringuet, 91128 Palaiseau, France.
§Present Address: LPNHE, Sorbonne Université, Université de Paris, CNRS/IN2P3, 75005 Paris, France.
‖Present Address: Hassan II University, Faculty of Sciences, Aïn Chock, BP 5366 Maarif, Casablanca 20100, Morocco.
¶Present Address: Université Grenoble Alpes, Université Savoie Mont Blanc, CNRS/IN2P3, LAPP, 74000 Annecy, France.
**Present Address: Institut de Physique Nucléaire de Lyon, CNRS/IN2P3, Université Lyon, Université Lyon 1, 69622 Villeurbanne, France.
††Present Address: Institut de Physique Nucléaire Orsay, CNRS/IN2P3, 15 rue Georges Clemenceau, 91406 Orsay, France.

## I. INTRODUCTION

Over the last two decades, measurements of neutrino oscillations in the three flavor framework have determined all mixing angles and mass splittings [1–3]. However, a recent reevaluation of the prediction of antineutrino spectra emitted by nuclear reactor cores revealed a 6.5% deficit between detected and expected fluxes at baselines less than 100 m [4,5]. This observed discrepancy, known as reactor antineutrino anomaly (RAA), with a significance of 2.7 standard deviations triggered a new set of reactor antineutrino experiments at very short baselines of about 10 m including the STEREO experiment.

Two primary hypotheses are mainly considered as explanation for the RAA. One is given by beyond-standard-model physics when assuming the existence of a sterile, i.e., nonweakly interacting, additional neutrino. If the mass splitting with respect to the known three active neutrino mass eigenstates is around 1 eV, significant oscillations toward this new neutrino eigenstate may be visible for reactor antineutrinos at baselines smaller than 100 m. As a result, a deficit in the detected absolute antineutrino flux is observable at those baselines. Such a sterile neutrino could also explain the lower than predicted electron neutrino rates measured in the calibration runs of the gallium solar neutrino experiments [6,7]. The antineutrino flux model, however, relies on single measurements of beta-spectra of thin foils of uranium and plutonium irradiated by thermal neutrons [8,9]. From those spectra, a nontrivial conversion to antineutrino spectra, yields the predicted antineutrino flux [4,10]. Erroneous steps in this procedure or the underestimation of its systematic uncertainties could also potentially explain the RAA or diminish its significance [11,12]. These might also explain a deviation between the predicted and the measured shape of the energy spectrum of reactor antineutrinos.

This article concentrates on an approach to test the sterile neutrino hypothesis independent of rate or spectral shape predictions as detailed in the following: The survival probability of initial antineutrinos at baselines below 100 m can be approximated in natural units as

$$P_{\nu_e \to \nu_e}(L, E) = 1 - \sin^2(2\theta_{ee})\sin^2\left(\Delta m_{41}^2 \frac{L}{4E}\right) \quad (1)$$

where $L$ and $E$ are the baseline and antineutrino energy, respectively, $\theta_{ee}$ denotes the mixing angle, and $\Delta m_{41}^2 = m_4^2 - m_1^2$ denotes the difference of the squared eigenvalues of the new mass eigenstate $m_4$ and the first mass eigenstate $m_1$. The original RAA is best explained by sterile neutrino oscillations with $[\sin^2(2\theta_{ee}) = 0.17, \Delta m_{41}^2 = 2.3 \text{ eV}^2]$ [13]. This point in the parameter space serves as benchmark in this article. Since the survival probability is dependent on the baseline $L$ and the antineutrino energy $E$, the existence of a sterile neutrino state manifests in a baseline and energy dependent distortion of the measured spectra. Thus, by performing measurements at multiple baselines, e.g., by segmenting a detector, a largely prediction-independent relative measurement can be performed.

The running very short baseline experiments DANSS [14], NEOS [15], PROSPECT [16], NEUTRINO-4 [17], and STEREO [18] report significant exclusions of the allowed parameter space established in Ref. [5]. Recent global analyses and some of the recently reported experimental results suggest the existence of a sterile neutrino at different parameter regions with significances up to about 3 standard deviations [17,19–22]. However, some of these proposed parameter regions are in conflict with other data including constraints from cosmology [23,24]. How these tensions will be resolved is yet unclear and no unambiguous measurement of sterile oscillations has been reported so far.

In this context, we present new results from the STEREO experiment. In 2018, STEREO reported the exclusion of a significant part of the allowed parameter space [18]. There, we also excluded the benchmark point at 97.5% C.L. using 66 days of reactor-on data. In the current article, we report improved results using 179 days of reactor-on data. In the following, all steps of the analysis chain are detailed. Several of these steps have been largely improved with respect to the previous version of the analysis presented in [18,25]. In Sec. II, we describe the experimental setup. Section III details the extended dataset, while Secs. IV and V discuss the prediction of the antineutrino spectrum of the reactor and the simulation framework of the STEREO experiment, respectively. After presenting the determination of the energy scale of the detector in Sec. VI, we give the event selection in Sec. VII, followed by a description of detailed corrections of our selection efficiency in Sec. VIII. Section IX illustrates experimental backgrounds in the STEREO detector. They are exploited in a discrimination technique regarding the time shape of scintillation pulses to extract antineutrino candidates in Sec. X. Finally, we present the current improved results of our oscillation analysis in Sec. XI before concluding in Sec. XII.

## II. EXPERIMENTAL SETUP

The STEREO experiment is located at the ILL research centre in Grenoble, France. Alongside the STEREO antineutrino detector, the ILL accommodates about 40 different instruments for neutron scattering, nuclear spectroscopy and fundamental physics with neutrons, taking advantage of the high neutron flux produced by its nuclear reactor RHF (réacteur à haut flux).

### A. The ILL reactor site

The RHF [26,27] is a heavy-water moderated reactor of 58.3 MW nominal thermal power. Nominal reactor cycles are 45 days long and alternate with shutdown periods for maintenance and exchanging the fuel element. There can be up to 4.5 cycles per year. The single fuel element consists of 280 curved aluminium plates enclosing the fuel, arranged in a hollow cylinder of 41 cm outer and 26 cm inner diameter by 81 cm height (for further details see [27,28]). Its compactness makes it suitable for searches of short baseline oscillations. The fuel is 93% enriched in $^{235}$U rendering the contribution of $^{239}$Pu to the total number of fissions negligible (cf. Sec. IV). The thermal power is monitored by observing the balance of enthalpy at the primary cooling circuit. Data taking of STEREO falls in a period of increased reactor maintenance, reducing the number of cycles to typically 2 to 3 per year. To increase the lengths of these cycles, the reactor was operated below nominal power. In general, the power lies between 50 MW and 56 MW and is constant for the respective cycle.

The STEREO detector is located on level C of the reactor building which is at the height of the fuel element. Its neighboring instruments D19 [29] and IN20 [30] extract intense neutron beams from the reactor moderator (cf. Fig. 1). They impose a background of rays of $\gamma$ and neutrons in varying rates on the STEREO detector, as well due to the $e^+$ energy loss and subsequent annihilation with an $e^-$ followed by a neutron capture on a Gd nucleus. The neutron thermalizes in the scintillator and is afterwards captured with a characteristic capture time of 16 μs. The average time of the full process including thermalization is 18 μs. The neutron capture is followed by a γ-cascade of about 8 MeV total energy. Due to the kinematics of the IBD reaction, the prompt signal carries the $\bar{\nu}_e$ energy information: $E_{e^+} = E_{\bar{\nu}_e} - \Delta M + m_e = E_{\bar{\nu}_e} - 0.782$ MeV where $\Delta M$ is the mass difference between the neutron and the proton and $m_e$ is the electron mass. The kinetic energy of the neutron is negligible.

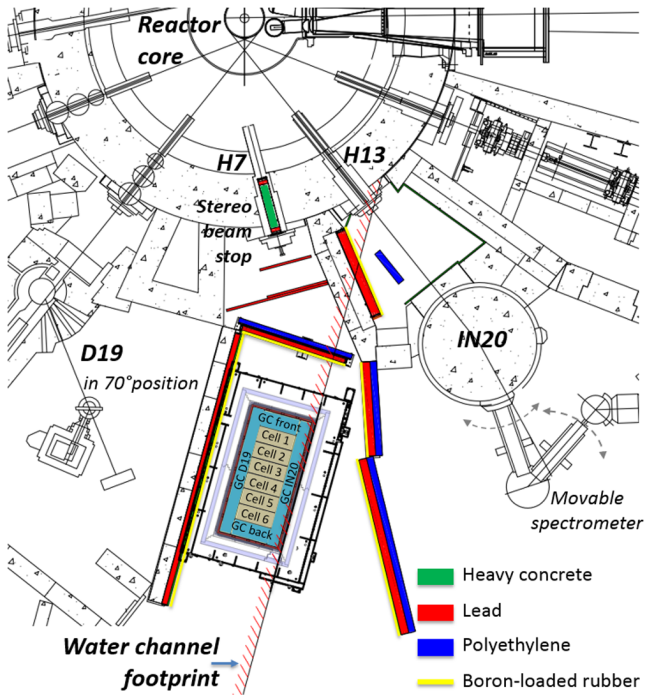

FIG. 1. Top view of the experimental site at level C of the ILL reactor building. Reinforced shielding of the STEREO site is highlighted. The shielding of the detector itself is shown in Fig. 2.

The segmentation of the fiducial volume, designated as "target" (TG), allows simultaneous spectrum measurements at six different baselines between 9.4 and 11.2 m (cf. Fig. 1). Each cell is 369 mm thick, 892 mm wide, and 918 mm high. The TG volume, delimited by an acrylic aquarium with 12 mm thick walls, is enclosed in a larger volume defining an outer crown around the TG, designated as "gamma-catcher" (GC) (cf. Fig. 2). The active volume of as stray magnetic fields of up to ∼1 mT [31] from high-field magnets (up to 15 T at the sample position inside the instruments). The shielding of the STEREO site was reinforced before the installation of the detector [25] and the STEREO detector is equipped with additional shielding as detailed in Sec. II B. The detector is partly covered by the transfer channel of the reactor (serving, for example, for the storage of used fuel elements). This channel is filled with light water and reduces cosmic-induced radiation at the STEREO site, yielding an inclination angle dependent shielding of about 15 metre water equivalent (m.w.e.) on average. The STEREO detector is aligned with this channel, resulting in an angle of $(17.9 \pm 0.2)°$ between the detector axis and the direction to the core. The center of the active detector volume is $(10.298 \pm 0.028)$ m away from the center of the reactor core, including a vertical distance of 0.21 m with the center of the active detector volume lying lower than the center of the reactor core.

## B. The STEREO detector

The STEREO detector uses the inverse beta decay (IBD) reaction $\bar{\nu}_e + p \rightarrow e^+ + n$ to detect $\bar{\nu}_e$ within a 1.6-ton gadolinium-loaded liquid scintillator target divided into six optically isolated cells [25]. The liquid scintillator is a LAB-based three-solvent-system using PPO and bis-MSB as fluors, while gadolinium (Gd) is dissolved in form of a Gd-beta-diketonate complex [32]. The IBD identification exploits the time structure of the IBD event: a prompt signal the GC has outer boundaries of $L \times w \times h = 2996$ mm $\times$ 1456 mm $\times$ 918 mm and consists of unloaded liquid scintillator of similar composition as the TG scintillator [32]. This crown allows to detect escaping $\gamma$-rays from events generated in the TG (511 keV $\gamma$-rays from the $e^+$ annihilation or $\gamma$-rays from the n-Gd cascade) and serves also as an active veto against external background. The GC is optically segmented in 4 cells, as illustrated in Fig. 2. The two cells (GC-Front and GC-Back) in front of the first and behind the sixth TG cell have the same geometry as the TG cells in order to reduce edge effects in the detector response. Separation walls within the same liquid are built with VM2000™ enclosed in an air-filled gap between two thin acrylic plates (2 mm). This particular design avoids a direct contact of the liquid with the VM2000™, which would otherwise lead to a decreased reflectivity at large angles of incidence [33,34]. Similarly, aquarium walls, side walls, and the bottom plate are made highly reflective by using the same design. Imperfect optical separations cause an optical cross-talk of several percent between cells, which is discussed in detail later.

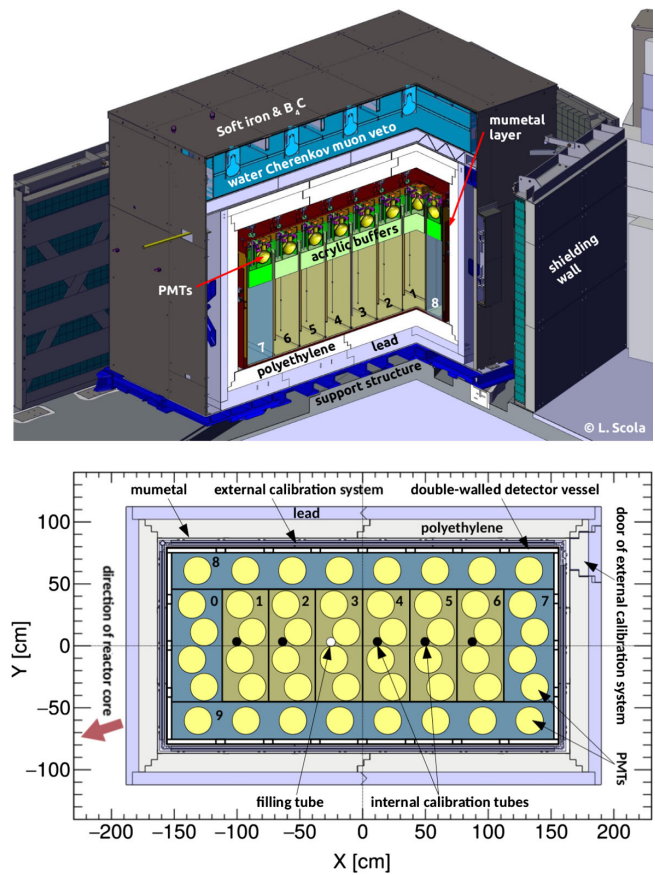

FIG. 2. Cutaway (top) and top view (bottom) of the STEREO detector setup. 1–6: Target cells (baselines from reactor core: 9.4–11.2 m); 0 and 7–9: Gamma-Catcher cells. The $Z$-axis starts at the bottom of the cells and points upwards, i.e., toward the muon veto. $B_4C$ layers on the inner shielding walls and the outer detector walls are not shown.

The scintillation light is read out by 48 8-inch photomultiplier tubes (PMTs), 4 for each TG cell and short GC cell and 8 for each long GC cell. They are located on top of each cell and separated from the scintillator by thick acrylics blocks (20 cm) designated as "buffers" (cf. Fig. 2). The optical coupling between PMTs and acrylic is provided by a bath of mineral oil [32].

PMT signals are continuously digitized at 250 mega samples per second using 14 bit ADCs. A dedicated electronic system [35] was designed for the STEREO experiment in order to trigger, process, and readout the PMT signals. Observation of a signal above the first level trigger threshold ($\sim 300$ keV energy deposition in one cell) triggers the processing of the event: a constant fraction discriminator (CFD) algorithm is used to determine the start time of the pulse of each individual PMT, not only the triggered ones. Then, the individual pulses are integrated to determine the amount of light detected by each PMT. The total and the tail pulse integral, $Q_{tot}$ and $Q_{tail}$, respectively, are obtained by a Riemann-integration over $N_{tot}$ samples (total pulse duration) and $N_{tail}$ samples (pulse tail duration). Typically 60 samples (240 ns) are enough to fully contain a pulse. $Q_{tot}$ is needed for the energy measurement, while $Q_{tail}$ allows to built a metric for particle identification by pulse shape discrimination (PSD), effectively separating interactions from strongly and lightly ionizing charged particles. Improvement of the PSD settings implemented in November 2017 resulted in a better separation between electronic (IBD-like) and proton (backgroundlike) recoil components as compared to Ref. [18]. In standard acquisition mode, only the processed data (CFD time, $Q_{tot}$ and $Q_{tail}$) are recorded for all channels. This allows for a very high instantaneous trigger rate and a negligible fraction of dead-time which is typically less than 0.02%.

The number of protons in the TG is derived from a measurement of the scintillator mass and its hydrogen fraction. The scintillator mass was determined during detector filling by comparing the masses of full and emptied scintillator barrels. The hydrogen fraction was determined by CHN element analysis (combustion analysis) of a scintillator sample [32]. The obtained value is $(1.090 \pm 0.011) \cdot 10^{29}$ TG protons. Its uncertainty is dominated by the precision of the combustion analysis. However, as the oscillation analysis presented in this article is not using a comparison with an absolute spectrum, the uncertainty of the proton number does not propagate to it. It is a cell-to-cell correlated uncertainty that becomes relevant for absolute rate measurements only.

In addition to the overall fiducial TG volume, it is also relevant to know the volume of individual cells. Exact cell dimensions and associated uncertainties could be measured for the empty detector before filling. From the CAST3M software package [36], which uses a finite element method, the deformation of TG cell walls due to few millimetres static difference in liquid levels was examined. The cell volumes were found to be 301.7 $\ell$ with a relative uncertainty of 0.83%. This uncertainty enters the oscillation analysis, as it is considered cell-to-cell uncorrelated (cf. Table II).

The detector light response and stability are continuously monitored. An LED-based light injection system is used to calibrate the PMTs at the photo-electron level and to monitor the linearity of the electronics. In addition, a set of radioactive $\gamma$-ray and neutron sources ($^{68}$Ge, $^{137}$Cs, $^{54}$Mn, $^{65}$Zn, $^{60}$Co, $^{42}$K, $^{24}$Na, $^{241}$Am/$^9$Be) is regularly deployed inside and all around the detector to monitor the detector response and to determine the energy scale. Sources can be deployed via three different calibration systems: (1) through vertical, Teflon-coated steel tubes spanning the full height of the TG, approximately at the center of each TG cell (except cell 3 where a filling tube made from pure Teflon is installed), (2) in a semiautomated positioning system along the perimeter of the detector, in-between the detector vessel and the shielding, (3) on a rail below the detector along its central long axis (cf. Fig. 2 and Table I).

In the first period of the data taking, called "phase-I" (cf. Sec. III), several deficiencies of the acrylics impacted the response of the detector. The most critical one was the loss of the oil-bath on top of two buffers (the one in the fourth TG cell and the one in the GC-front cell) due to leaks in the buffer aquariums. This resulted in a drop of the collected light for these cells by a factor 2.5 compared to the other cells. In addition, most of the optically separating walls lost their tightness allowing the LS to fill the air gap around the VM2000™ layer. As a consequence, the optical cross-talk between cells increased typically from a value of 5% to 15%. In the second phase of data taking starting mid-2017 ("phase-II", cf. Sec. III), full light collection was restored and light cross-talk was significantly reduced after maintenance on the acrylic walls and buffers. Optical cross-talk between TG cells is now close to the design value with an average value of 6.9% (9.6% for cell 3 to 4) and is stable within ±7% (relative deviations) over 16 months, except for cross-talk between cell 3 to 4 which has increased from a value of 6.4% to 10.2%. The aquarium walls between TG and GC could not be repaired, so the average value from the first and sixth TG cell to the neighboring short GC cell is 10.5% and the average value from each TG cell to each long GC cell is 2%.

TABLE I. Coordinates of source deployment positions in the internal calibration tubes (top), selected external calibration positions along the $X$-axis (middle), and selected underneath calibration positions along the $X$-axis (bottom). The vertical distances of the internal positions to the bottom of the tubes is given as $\Delta Z$. The bottom of each tube has a distance of 2.5 cm to the bottom of the cells. Coordinates marked with an asterisk (∗) can be chosen freely, the others are given by the detector geometry. See Fig. 2 for the definition of the cell numbers and coordinate system.

| Internal calibration tubes | | | | | |
|---|---|---|---|---|---|
| Cell | $X$/ cm | $Y$/ cm | | Z-position | $\Delta Z^*$/ cm |
| 1 | −100 | 3 | | Top | 80 |
| 2 | −63 | 3 | | Mid-top | 60 |
| 3 | | | | Middle | 45 |
| 4 | 12 | 3 | | Mid-bottom | 30 |
| 5 | 49 | 3 | | Bottom | 10 |
| 6 | 87 | 3 | | | |

| External calibration system | | | | | |
|---|---|---|---|---|---|
| Cell | $X^*$/ cm | $Y$/ cm | | Z-position | $Z^*$/ cm |
| 1 | −93 | ±82 | | Top | 87 |
| 2 | −56 | ±82 | | Middle | 45 |
| 3 | −19 | ±82 | | Bottom | 17 |
| 4 | 19 | ±82 | | | |
| 5 | 56 | ±82 | | | |
| 6 | 93 | ±82 | | | |

| Underneath calibration system | | |
|---|---|---|
| Cell | $X^*$/ cm | $Y$/ cm | $Z$/ cm |
| 1 | −93 | 3 | −10 |
| 2 | −56 | 3 | −10 |
| 3 | −19 | 3 | −10 |
| 4 | 19 | 3 | −10 |
| 5 | 56 | 3 | −10 |
| 6 | 93 | 3 | −10 |

To reduce $\gamma$-ray and neutron background, the STEREO detector is enclosed in a passive shielding of about 65 tons (cf. Fig. 2) composed of borated polyethylene (29.7 cm on top, 14.7 cm on sides, and 20 cm below) and lead (15 cm on top, 10 cm on sides, and 20 cm below). A water Cherenkov detector used as an active muon veto is placed on top of the lead shielding. It shows a detection efficiency for vertical muons greater than 99.5%, with a very good stability: observed variations are up to 0.4% in phase-I and, after the exchange of failing PMT bases, stay within 0.1% in phase-II. Overall, cuts on the veto (cf. Sec. VII) reduce correlated background by about 30% (limited by the geometric coverage of the veto). A magnetic shielding composed of several layers of different materials protects the STEREO system against the effect of the magnetic field generated near the STEREO site. From outside to inside, a soft iron layer surrounds the lead shielding and the muon veto detector, a mu-metal layer is inserted between the polyethylene and the STEREO detector (cf. Fig. 2) and cylinders of mu-metal are placed around the PMTs. The soft iron layer is covered by boron-loaded rubber (containing $B_4C$ powder) absorbing the ambient thermal neutrons present in the reactor hall.

## III. DATASET

The data collection considered here includes a total running time of 565 days [37–43]. As seen in Fig. 3, reactor-on and reactor-off phases alternate, allowing for measurements of background pure samples roughly about every two months. The first 24 h following a large transient of the reactor power are removed from the dataset to mitigate the corrections to be applied to the prediction of the spectrum shape (see Sec. IV).

In March 2017, the STEREO detector had to be retracted from its measuring position to allow for maintenance work at the reactor. In particular, the through-going beam tube H6/H7 including its STEREO beam-stop was removed (cf. Figs. 1 and 5). The STEREO collaboration took advantage of this period to repair defective acrylics and perform maintenance work on the detector. As light collection, cross-talk, and potentially reactor background conditions have changed, data recorded before and after the repair are treated separately in the analysis. The dataset until March 2017 is labelled "phase-I" while the dataset recorded after the recommissioning is labeled "phase-II".

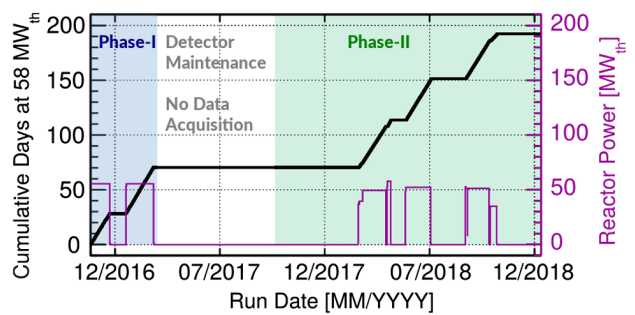

FIG. 3. Cumulative (black) and daily (purple) reactor power over calendar day. Alternating reactor-on and reactor-off phases as well as a long reactor shut-down phase used for detector maintenance are depicted.

As mentioned in Sec. II B, a change in the PSD settings occurred during the first reactor-off period of phase-II. To be consistent with reactor-on data, data recorded before the change (37 days) are removed from the dataset of phase-II, but retained for cross-checks between the two phases. This yields a running time of 528 days, of which 91% are used in the oscillation analysis. The remainder comprises calibration runs (3%), runs during or following a large transient of the reactor power (3%), idle time in-between runs (2%, dominated by time in-between calibration runs and after unexpected stops of the data acquisition), and test runs (1%). Taking into account the veto-time induced by the event selection (cf. Sec. VII), this results in a total live-time of 118.5 (60.8) days of reactor-on data and 211.2 (24.0) days of reactor-off data in phase-II (phase-I).

During most long reactor shutdowns, the water level in the reactor pool above the core is reduced from its nominal value of 15 m to 7 m, resulting in an increase of the detected muons in STEREO of about 2.5%. To study this effect on IBD candidates, the reactor-off dataset includes 61 (21) days of data recorded with a full pool in phase-II (phase-I). In contrast to the pool, the water level in the channel above the detector did not change (cf. Fig. 1).

## IV. REACTOR PREDICTION

In the oscillation analysis described in Sec. XI, the nonoscillated spectrum shape common to all cells is let free in the fit, suppressing the sensitivity to the initial prediction of the antineutrino spectrum emitted by the reactor. However, a precise estimate of the expected spectrum lays the basis for absolute measurements, to be discussed in future publications. The evolution of the fuel content of the ILL core during a cycle at nominal full power was first simulated using the FISPACT code [44]. Due to the high enrichment in $^{235}$U of the fuel (93%) very little plutonium is produced during an ILL-cycle, the mean fission fraction of $^{239}$Pu was found to be 0.7% only. This contribution has virtually no impact on the predicted emitted antineutrino spectrum. Therefore, the Huber model of a pure $^{235}$U spectrum [10] is used. Few corrections must be applied to predict the spectrum expected in the STEREO detector [45].

The Huber model provides a snapshot of the fission spectrum after only 12 h of irradiation of a $^{235}$U target. Thus, the accumulation of the fission products with lifetime comparable to or larger than 12 h must be estimated to extrapolate to a typical 50-day long ILL cycle. This effect is called the "off-equilibrium" correction. After a reactor stop, the spent fuel is stored at specific locations in the water channel above the STEREO detector. The same long-lived fission products present in the spent fuel keep emitting antineutrinos that contribute to the spectra measured during the reactor-on and reactor-off periods. These two corrections have been calculated by coupling the isotopic inventory predicted by FISPACT along the reactor history with the BESTIOLE code [46] providing the antineutrino spectrum of each isotope. They are displayed in Fig. 4 as the mean relative correction factor $\delta$ to be added to the Huber model of the antineutrino energy spectrum for a 50-day long cycle at nominal power:

$$\text{Huber}(E_\nu) \rightarrow \text{Huber}(E_\nu) \times [1 + \delta(E_\nu)]. \quad (2)$$

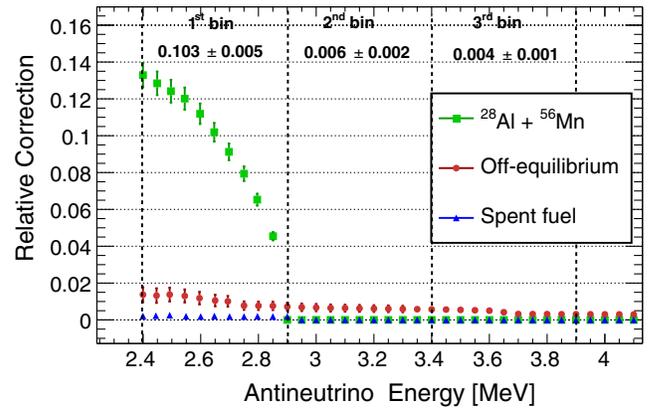

FIG. 4. Relative corrections to be added to the Huber model prediction for antineutrinos at the STEREO detector position: off-equilibrium effect (circles), residual antineutrinos from spent fuel (triangles) and contribution of the $\beta^-$-decays of $^{28}$Al and $^{56}$Mn produced by neutron captures in the reactor (squares). The shown bin-to-bin correlated uncertainties include tolerances in the structural elements as well as the amount of $^{55}$Mn in the aluminium alloys. The dotted vertical lines represent the first energy bins of the oscillation analysis. Their associated mean corrections are given per bin, as well.

As expected, the corrections affect the lowest energy part of the antineutrino spectrum. The visible structure in energy is attributed to dominant long-lived fission products $^{144}$Pr, $^{92}$Y, and $^{106}$Rh with endpoints reaching 3 to 3.5 MeV along their respective decay chains. Their amplitude is smaller than typical off-equilibrium and spent fuel effects computed for commercial reactors due to the short duration of the ILL cycles, which prevent significant accumulation of very long-lived isotopes. In the STEREO analysis, the prompt energy threshold (equivalent to $E_\nu > 2.4$ MeV) and the removal of the first 24 h of data following a large transient of reactor power further mitigate the corrections.

The last and dominant correction also shown in Fig. 4 is linked to the very high neutron density in the core and the large amount of aluminium in the fuel element itself and the mechanics around it (beam tubes and heavy water vessel). Complementary MCNP-2.5 [47] and TRIPOLI [48] simulations were used to infer the neutron spectrum in the various structural elements and compute their activation rate and their subsequent $\beta^-$-decays. The detailed calculation with TRIPOLI, including all the geometry of the core (see Fig. 5), shows the dominant activation of $^{27}\text{Al}$ by neutron capture. It contributes to about 26% of all $(n,\gamma)$ reactions and absorbs $(0.3426 \pm 0.0004)$ neutrons per fission. This uncertainty is given by the simulation with a control of the neutron balance at the 0.1% level. It does not include uncertainties in the amount of aluminium. The source volume effect was taken into account by calculating the geometrical solid angle to the detector from each emission point. Compared to the source volume effect of antineutrinos produced by fission, the correction amounts to $(+1.0 \pm 0.3)$% leading to an effective $^{27}\text{Al}$ capture rate of $(0.346 \pm 0.001)$ per fission. The $\beta^-$-decay of $^{28}\text{Al}$ produces antineutrinos with energies as high as 2.86 MeV leading to a mean correction of $-10.0$% in the first 500 keV energy bin of the oscillation analysis. The next contribution, yielding a $-1.0$% correction, is from the capture on $^{55}\text{Mn}$. We attribute a 5% uncertainty to the total correction, accounting for tolerances in the structural elements as well as in the amount of manganese in the aluminium alloys. All other $(n,\gamma)$ reactions lead to either stable nuclei or antineutrinos below threshold.

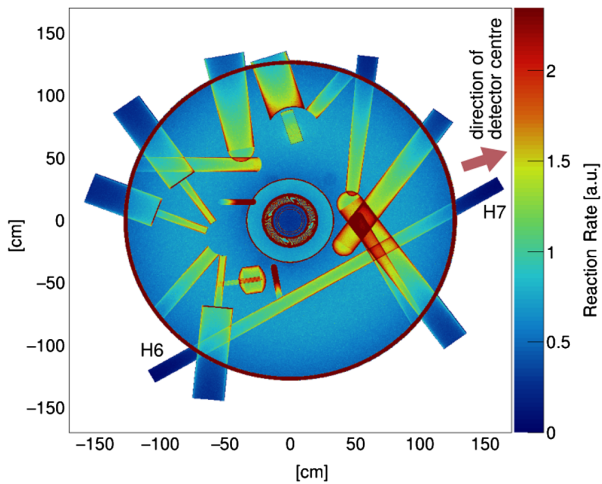

FIG. 5. Top view of the $^{27}\text{Al}(n,\gamma)^{28}\text{Al}$ reaction rate in the reactor heavy water tank, integrated over the 340 cm height, for STEREO phase-I. During STEREO phase-II, the through-going beam tube H6/H7 was absent (taken into account in the correction shown in Fig. 4). Beam tubes made of Zircaloy-4 contain negligible amounts of Al and are not visible in the figure. Courtesy of A. Onillon, CEA.

The initial prediction of the spectrum shape used in the following is then taken from the simulation of about 10 million IBD events weighted according to the Huber model for pure $^{235}\text{U}$. It is corrected for the three abovementioned effects. Since the antineutrinos from $^{28}\text{Al}$ and $^{56}\text{Mn}$ have a similar origin in space as those from the core (including off-equilibrium antineutrinos) and since the antineutrinos from spent fuel are negligible, we apply the oscillation model to the full corrected spectrum.

## V. MONTE CARLO SIMULATIONS

The Monte Carlo simulation framework of STEREO is built using the GEANT4 toolkit (version 10.2) [49,50], and it is based on generic libraries for liquid scintillator detectors GLG4Sim, tested and developed in experiments like KamLAND and Double Chooz [51,52]. The simulation includes all aspects of an event from particle interactions in the detector to light collection in the PMTs. It is implemented such that MC output has the same format as real data. At this point, real data and simulation can be processed analogously during the analysis. For this purpose, the entire geometry of the detector has been implemented, including all the mechanical parts, photomultipliers, shielding components, the Cherenkov-veto, the calibration systems, and the main reactor building elements. The latter is mostly relevant for the simulation of the cosmic background, making use of the CRY libraries [53].

Special interest has been put into replicating the scintillation process in the simulation. Several intrinsic properties of the liquid scintillator have been tested in dedicated laboratory measurements [32]. These, together with information from calibration data, have been used to fine-tune the scintillator parameters in the simulation of the STEREO experiment. Some of these variables are the quantum yield of the fluors, the total attenuation length and light yields of both liquid scintillators. The quantum efficiency of the PMTs has also been accurately described in the simulation using measurements from Ref. [54].

In the simulation, the TG proton number is calculated from the vessel volumes, the densities of the solvents and the chemical composition of all the scintillator components. The linear alkyl benzene (LAB) as main component (73 wt.%) does not have a well-defined molecular formula, since it is a mixture with hydrocarbon chains of different lengths. Therefore, approximated values are used for the number of hydrogen and carbon atoms per molecule in the MC to match the average molecular mass as specified by the supplier. The scintillator density is calculated from the individual densities of the solvents which could also introduce a slight bias in the TG proton number of the MC. This is mainly due to the small contributions of the wavelength shifters and the Gd-complex, which are not taken into account. Overall, the calculated TG proton number in the simulation is less accurate than the one obtained from the mass and hydrogen fraction measurements as explained in Sec. II B. Therefore, a correction factor of $0.983 \pm 0.010$ on the normalization of the simulated antineutrino events is applied for the scintillator below the acrylic buffers (height 91.8 cm). For the small volumes at the side of the acrylic buffers, average values of the liquid levels in phase-I and phase-II are used. The associated uncertainty related to this subvolume can be neglected. Since, in Sec. VII, we find the fraction of selected IBD candidate events outside of the TG volume to be much smaller than 1%, we can neglect volumes other than the TG in the calculation of the proton number.

The bulk properties of reflectivity and transmission of the separative plates were also carefully treated. They directly impact the collection of light, the top-bottom asymmetries, and thus the energy resolution. As mentioned earlier, the walls are made of two acrylic plates with, enclosed between them, one or two sheets of VM2000™ mirror films and a nylon mesh that ensures an air gap between the plates and the films. VM2000™ mirrors (current name is Enhanced Specular Reflector—ESR) are multilayer films made of highly birefringent polymers and were developed by 3M [55]. They exhibit a high reflectivity in a large range of wavelengths at any incident angle in air. The reflectivity of the film used for STEREO was measured with a spectrophotometer to be above 98% in the 400–950 nm range in air, at an angle of incidence of 7°. As stated in Sec. II B, liquid scintillator entering and filling the air gap of the separation walls degraded the reflective capabilities of the ESR film, resulting in increased light cross-talks. Indeed, the strong decrease in reflectivity of an ESR film in a liquid is expected at large angles [33,34]. At 450 nm, the transmission was, e.g., measured with a spectrophotometer to increase up to 50% for angles of incidence larger than 75°, when the VM2000™ was placed in liquid scintillator. To account for this scenario, a versatile description of light cross-talk between cells was implemented in the simulation of STEREO via the introduction of individual filling levels for all acrylic walls (side walls, separation walls, bottom plate) and a model describing the effect of the liquid on the reflectivity of the ESR film. The addition of absorption effects were found to be necessary to accurately describe the widths of the calibration peaks and to reduce the differences of top-bottom asymmetries between data and MC. They were implemented via two terms: absorption of part of the light transmitted through the ESR films due to the shadow of the mesh, and reduced reflectivity of the Teflon-coated calibration tubes (considered as perfect diffusers in earlier analyses). A separate tuning of these two degenerated parameters was made possible by the fact that cell 3 has no calibration tube, but instead a filling tube, made of highly reflective bulk white Teflon.

The main parameters of the model (levels of liquid in the wall, absorption terms, transmission at large angle of incidence in VM2000™) were optimized in a global approach to reproduce the measured light cross-talks and charges collected by the PMTs for sources at different heights. The single $\gamma$-peaks of the $^{24}$Na and the entire spectrum of the $^{54}$Mn calibration sources deployed at five different positions within each of the calibration tubes were used for this purpose. The well-localized vertices of the 834 keV $\gamma$-rays from $^{54}$Mn decays and the coincident $\gamma$-rays at 1368 and 2754 keV from $^{24}$Na decays provide complementary information. Examples of agreement on the collected charge between data and simulation are shown in Fig. 6, when the $^{24}$Na source is placed inside the considered cell or in a neighboring cell. The collection of light is well reproduced in both cases. When the source is in a neighboring cell, the prominent structure at low charge was used to gain sensitivity to light cross-talk and to set the liquid level parameter for the considered separating wall. In general, better agreement is observed for the middle source positions rather than for top or bottom positions.

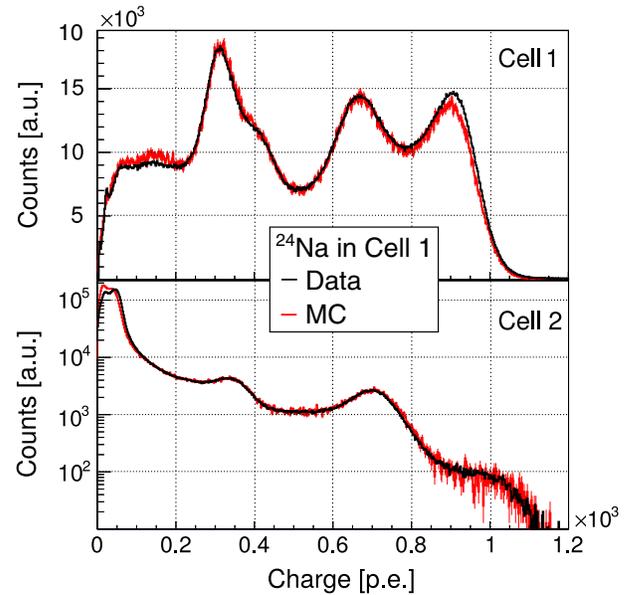

FIG. 6. Measured (black) and simulated (red) collected charges in cell 1 (top) and in cell 2 (bottom) when the $^{24}$Na source is located at the middle of cell 1.

To quantify the residual discrepancies between simulated and measured spectra, homogeneous scaling factors, denoted as dilatation factors, are determined for the simulated charges such that each simulated spectrum would best match its corresponding measured spectrum in a $\chi^2$-fit. Figure 7 shows an overview of the dilatation factors obtained for each of the 25 measured positions of the manganese calibration source (cell 1, 2, 4, 5, 6 and height 10, 30, 45, 60, 80 cm from the bottom of the calibration tubes, cf. Table I). Most factors are within a $\pm 1\%$ band around 1 illustrating the very good agreement at the raw charge level, before calibrating any energy reconstruction. Two outliers at the 2%–3% level correspond to the 10 cm position in cells 4 and 5. The origin of these could be due to a local effect of the bottom plate in these cells, to be investigated in a future version of the simulation. This fine tuning of the optical parameters was performed by comparing the simulation with the data of 28th April 2018, corresponding to about the middle of phase-II. It is the purpose of the energy reconstruction to correct, to first order, the residual time evolution of the light collection in the detector, mitigated during phase-II (see Sec. VI).

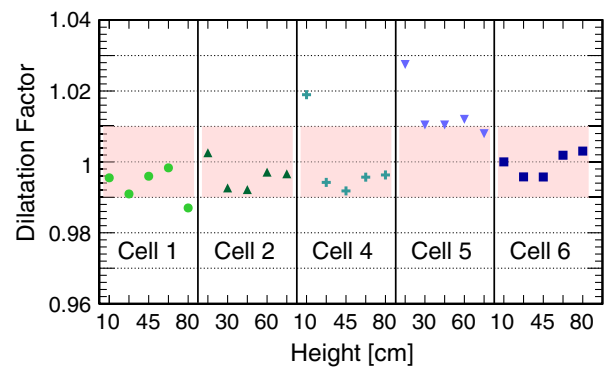

FIG. 7. Summary of all the dilatation factors fitted from the 25 measured and simulated positions of the Mn source. A ±1% band containing most factors is highlighted.

Lastly, additional attention was given to the simulation of neutron mobility and gammas released by their capture by Gd nuclei. In particular, the refinement of the description of thermal scattering by including molecular data for hydrogen [56] improved the agreement between the simulated and observed capture times using an AmBe source. Moreover, the description of the deexcitation cascades of the relevant Gd isotopes $^{156}$Gd and $^{158}$Gd were significantly improved, after making use of the FIFRELIN code [57]. Since the Gd level schemes are not experimentally known up to the energies of the relevant excited states after thermal neutron captures, models are required to describe the de-excitations of the formed nuclei. In the FIFRELIN simulation, a set of nuclear level schemes is sampled from the most updated experimental data and the recommended nuclear models and data evaluations. Deexcitation processes are then generated within a Monte-Carlo Hauser-Feshbach framework. This approach allows to take into account nuclear structure uncertainties, and provides an accurate description of the energy and multiplicity distributions of the de-excitation products. An high-statistic sample of de-excitation products generated from FIFRELIN is used in the STEREO simulation on an event-by-event basis for each cascade following a neutron capture on the $^{155}$Gd or $^{157}$Gd isotopes [58,59].

# VI. DETECTOR RESPONSE

A good understanding of the energy scale is an essential prerequisite for the analysis presented in this article. Because of the steep variations of the detected IBD spectrum the study of its shape is extremely sensitive to the control of the energy scale [60]. Based on the comparison between identical cells, the search for the oscillation signal presented here mitigates the impact of an overall bias at the detector level, but the uncorrelated effects between cells must be accurately estimated. The strategy for studying the detector response for phase-II of the experiment remains similar to that described in Ref. [25] for phase-I. However, between the two phases the detector was emptied and the acrylics repaired. By this, the light collection properties including the light cross-talk between cells, a dominating effect in the description of the detector response of phase-I, have changed substantially. Moreover, the Monte Carlo was retuned to new optical parameters. The systematic uncertainties of the energy reconstruction are therefore considered to be essentially independent between the two phases.

## A. Nonlinearities

The detector response deviates from a perfect linear model due to two main causes. Quenching for high $dE/dx$, intrinsic to the liquid scintillator, leads to a nonlinear response for low energy deposits. This phenomenon is described by an effective Birks coefficient $k_B$ [61]. Moreover, Cherenkov-radiation contributes to the nonlinearity.

In addition to the nonlinearities, there is a complex collection of light in the STEREO detector: the attenuation length in the liquid induces a dependence on the vertex and, although stabilized, light cross-talks are still present between cells.

To disentangle the various effects, pointlike radioactive $\gamma$-sources ($^{137}$Cs, $^{54}$Mn, $^{65}$Zn, $^{42}$K, $^{60}$Co, $^{24}$Na) and the $\gamma$-neutron source $^{241}$Am/$^9$Be are deployed in the internal calibration tubes and the external calibration system at all Z-positions (cf. Table I) to study the quenching effect. These measurements are performed during periods where the reactor is turned off to minimize environmental background, in particular from $^{41}$Ar present in the reactor building. The light cross-talks are estimated by looking at the ratio of the charge collected in a neighbor cell to the charge collected in the calibration cell. This analysis is performed using $^{54}$Mn source runs. These cross-talk values are then used to select full energy deposits in the calibration cell by limiting the charge detected in the neighbor cells. The cell studied is taken as the one next to the source cell to have a better separation of the $\gamma$-ray in the case of multi-$\gamma$ sources. In addition, for multi-$\gamma$ sources, a deposited energy, corresponding to the second $\gamma$-ray, is required in the gate cell, which is opposite to the calibration cell. The light cross-talks between the gate cell and the calibration cell are negligible. The same method is used to separate the neutron and the $\gamma$-ray of the AmBe source.

Identical cuts are applied to the measured and simulated events. The events with a charge included in a determined range around the peak positions are selected. For measured and simulated events, the mean deposited charge is taken as the arithmetic average of the charges of these events. The mean true deposited energy, available in simulations only, is also determined by an arithmetic average over the selected events. The calibration coefficient for data and for MC is calculated for each source as the ratio of the respective mean charge over the mean deposited true energy. To compare the results in different cells, the coefficients are normalized to the one of the $^{54}$Mn source, chosen as the calibration anchor point of the experiment (see Sec. VI B). As the normalized coefficients agree within their uncertainties across cells, it is then possible to average them over the 6 cells. For the special case of the AmBe 4.4 MeV $\gamma$-ray, the average is taken only over 4 cells. This is due to the fact that the calibration tubes are not centered in the cells along the $X$-axis (cf. Fig. 2). Therefore, in case the tube in, e.g., cell 5 is used, the number of neutrons reaching the further cell, i.e., cell 6, is too little. Thus, only the closer cell, i.e., cell 4, can be used as gate cell. This prevents calibration if cells 1 and 4 are the cells of interest, because the neighbor cells 0 and 3 do not have a calibration tube (cf. Fig. 2). The uncertainty takes into account the discrepancies obtained by varying the cuts on the deposited charge in the neighbor cells and the limits of the averaging range of the coefficients.

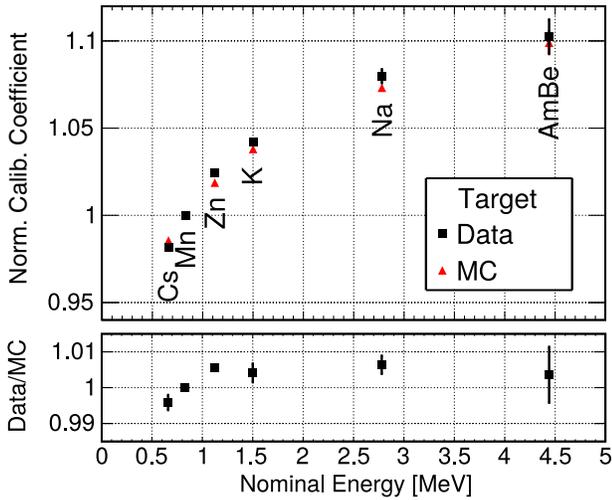

FIG. 8. Top: calibration coefficients for various radioactive sources at midheight, normalized to the $^{54}$Mn point and averaged over all TG cells. The true deposited energy corresponding to the charge peak value is reported on the horizontal axis. For $^{24}$Na, only the higher energy $\gamma$-ray is used. Bottom: ratio of data and MC curves, after tuning of the Birks coefficient $k_B$.

The averaged coefficients in Fig. 8 clearly show the expected quenching effect, responsible for a reduced light yield at low energy. After adjusting the effective Birks coefficient in the MC, the agreement between experimental and simulated curves reaches sub-% accuracy with $k_B = (0.096 \pm 0.007)$ mm/MeV. Note that the quenching effect associated to positrons (prompt IBD signal) is smaller than for $\gamma$-rays. However, a good control of the reconstructed $\gamma$-energies is mandatory to study the systematic effects of the energy scale with radioactive sources and to accurately control the neutron detection efficiency via the $\gamma$-cascades of Gd isotopes.

## B. Calibration anchor point

As described in Ref. [25], conversion factors between collected PMT charge and light cross-talk between neighboring cells on the one hand, and reconstructed energy on the other hand, are calculated by iteratively tuning these parameters to $^{54}$Mn data, taken on a weekly basis. This source emits a single 835 keV $\gamma$-ray with an interaction length of $\sim$10 cm in the liquid scintillator, comparable to the dimensions of a cell along the $X$-axis ($\sim$37 cm), but significantly smaller than the dimensions along $Y$ and $Z$ (both $\sim$90 cm). Thus, events attributed to a specific TG cell are mainly acquired when the $^{54}$Mn source is located inside the same cell. However, a few events are also recorded when the source is located in a neighboring TG cell. For the special case of cell 3, only events from the neighboring cells are recorded (cf. Fig. 2).

A set of parameters representative of the average response in the entire volume of the cell, the only quantity accessible with the formalism of energy reconstruction, is obtained by taking the average of the responses of five source positions distributed along the vertical calibration tube ($Z$-direction, cf. Table I). As for phase-I, the experimental and simulated $^{54}$Mn peaks can be aligned simultaneously in all cells with high precision (cf. Fig. 9). The residual cell-to-cell fluctuations are reduced to the 0.2% level and enter the oscillation analysis (cf. Table II).

The $^{54}$Mn source was chosen to define the reference $\gamma$-energy, because among the sources with long half-life, it is the single-$\gamma$ source which has the highest $\gamma$-energy being reliably separable from $\gamma$-rays of $^{41}$Ar decays (1294 keV). As $^{41}$Ar concentrations in the reactor building vary during reactor-on periods, single-$\gamma$ sources with higher energies, like $^{65}$Zn (1116 keV), do not constitute an applicable alternative. The choice of a reference $\gamma$-energy at 835 keV corresponds to a larger quenching effect than for the positrons of the IBD candidates, in the [1.625, 7.125] MeV range. Therefore, the reconstructed energies in that range are a few percent higher than the "true" values. There is no attempt to correct the data for this effect. Instead, all relevant effects are included in the MC and any model or true energy information to be compared with the data is forward folded with the simulated detector response.

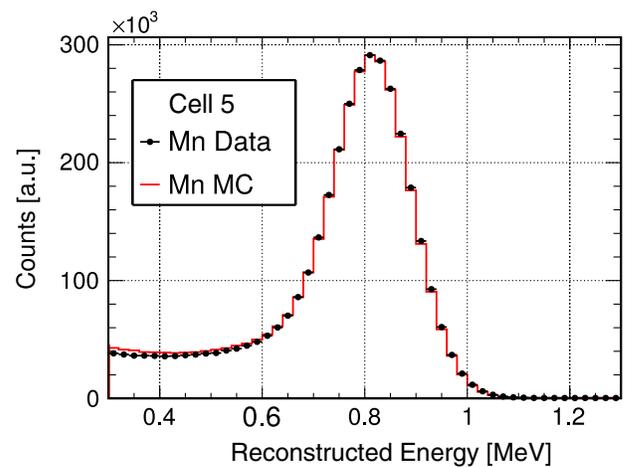

FIG. 9. Comparison of measured and simulated spectra in cell 5 for the average of the 5 vertical deployment positions of a $^{54}$Mn source in cell 5 (cf. Table I).

TABLE II.   Overview of relative systematic uncertainties entering the oscillation analysis (Sec. XI). Cell-to-cell correlated normalization parameters are not listed, since the oscillation analysis is insensitive to common shifts among detector cells.

| Type | Relative uncertainty | Reference |
|------|---------------------|-----------|
| Normalization (uncorrelated) | | |
| Cell volume | 0.83% | Sec. II B |
| Neutron efficiency correction | 0.84% | Sec. VIII |
| Energy scale (uncorrelated) | | |
| Mn anchor point | 0.2% | Sec. VI B |
| Cell-to-cell deviations | 1.0% | Sec. VI F |
| Energy scale (correlated) | | |
| Time stability | 0.3% | Sec. VI D |

This avoids a troublesome unfolding procedure and guarantees the most accurate treatment of the data.

### C. Energy resolution and spatial nonuniformity

Once the parameters of the energy reconstruction are fitted on the $^{54}$Mn data and the $k_B$ coefficient is optimized, different radioactive sources are studied in order to verify the good quality of the agreement between data and MC over the widest possible range of energy and position. For each nominal energy, the reconstructed energy $E_{rec}$ and the resolution, defined as the ratio $\sigma(E_{rec})/E_{rec}$, with $\sigma(E_{rec})$ the standard deviation to the right of the full energy peak, are reported. Figure 10 shows that the mean ratio $E_{rec}^{Data}/E_{rec}^{MC}$ over the five vertical positions stays within $\pm 1\%$ for all sources. The resolution, shown in Fig. 11, follows a $1/\sqrt{E}$ law at low energy and then saturates around 5% at higher energy. This behavior is reproduced by the simulation and is interpreted as the convolution of the pure statistical resolution with the variation of a few per cent of the collected light according to the altitude of the energy deposit in the cell.

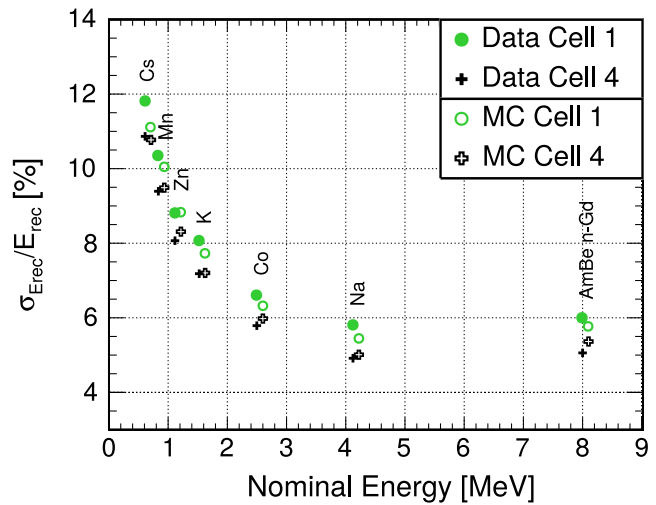

FIG. 11.   Energy resolution as a function of the nominal energy of emitted $\gamma$-rays for the average of the 5 vertical deployment positions of the sources (cf. Table I). Filled (open) markers are for data (MC). When two $\gamma$-rays are emitted, the energy used here is the sum. Cell 4 is representative for the four inner TG cells, while cell 1 is representative for the two outer TG cells.

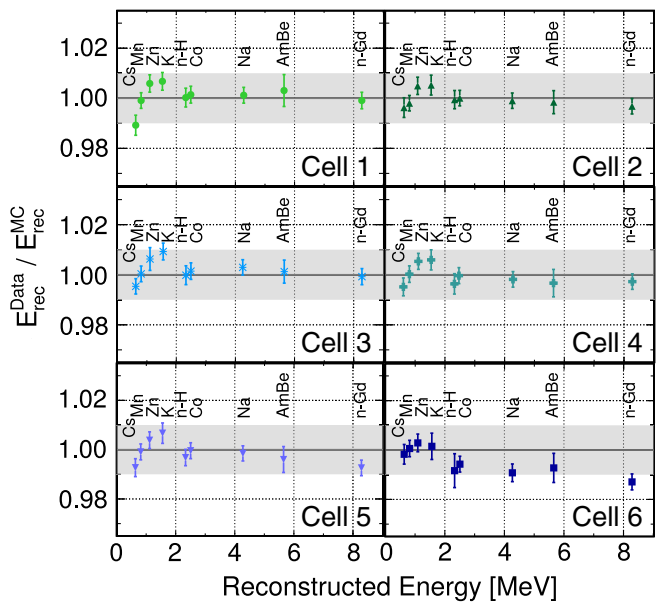

FIG. 10.   Ratio of reconstructed energy between data and MC as a function of the reconstructed energy of emitted $\gamma$-rays for the average of the 5 vertical deployment positions of the sources (cf. Table I). When two $\gamma$-rays are emitted, the energy used here is the sum. Data points are all contained in an uncertainty band of $\pm 1.0\%$ shown in grey. Uncertainties include effects from data statistics, as well as fit range and time dependence of the peaks, but no correlations across sources are assumed.

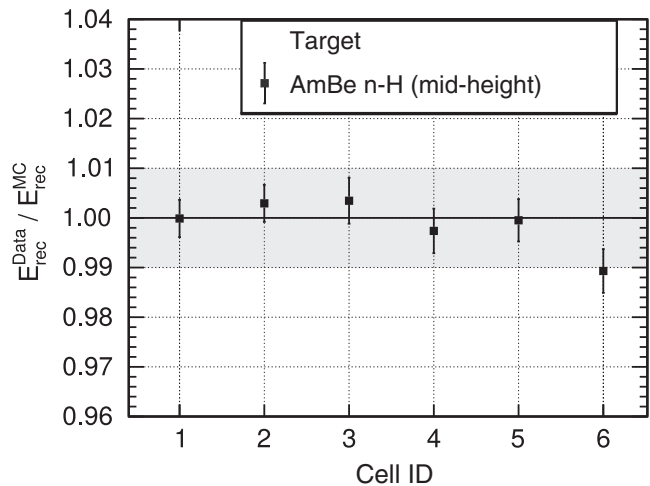

FIG. 12.   Ratio of reconstructed energy between data and MC of neutron captures by hydrogen for the average of 3 horizontal deployment positions of an AmBe source at midheight (2 external and 1 internal $Y$-positions, cf. Table I). Data points are all contained in an uncertainty band of $\pm 1\%$ shown in grey.

The deployment of γ-sources in the internal calibration tubes only probes the detector response in the central region of each cell. At the cell border ($Y$-direction), additional effects might influence the detector response. To probe the border regions of the six cells, an AmBe source is deployed outside the detector vessel, but inside the shielding, via the external calibration system (cf. Fig. 2). Per cell, six positions (two $Y$-positions and three $Z$-positions, cf. Table I) are investigated. All points are located at the center of the cell ($X$-axis). The AmBe source emits neutrons which are dominantly captured by hydrogen, before reaching the TG. The agreement between data and MC for both sides and the center of the detector is at the 1%-level across all cells (cf. Fig. 12). Since we find the same values when comparing the external calibration points with the internal points, we do not consider any additional correction or uncertainty for the reconstructed energy of events close to the cell borders.

### D. Time stability

To complete the study of systematic uncertainties, a set of events spatially distributed as similarly as possible to IBDs is ideally sought. A first sample of useful data is provided by the capture of neutrons from cosmic rays by hydrogen. The simulation of cosmic rays in the STEREO environment confirms the predominant role of the lead shielding in the generation of secondary neutrons. However, a significant fraction of these neutrons are absorbed by the materials arranged around the TG volume for this purpose: borated polyethylene, the GC and acrylic buffers. Remaining neutrons are those with high energy. Their capture vertices are widely distributed in the whole detector. The n-H capture peak is isolated in the data by looking for events around 2.2 MeV less than 300 μs after a muon signal in the veto or the detector. The central value and standard deviation of the reconstructed energy peak are extracted from the Gaussian core portion of a Crystal Ball fit function. Defining 20-day time bins, the detector response is found to be very stable at the 0.3% level in each cell (cf. Fig. 13). The remaining drifts are fully correlated across all cells. Therefore, this uncertainty on the time stability is taken as estimate of the cell-to-cell correlated systematic uncertainty (cf. Table II).

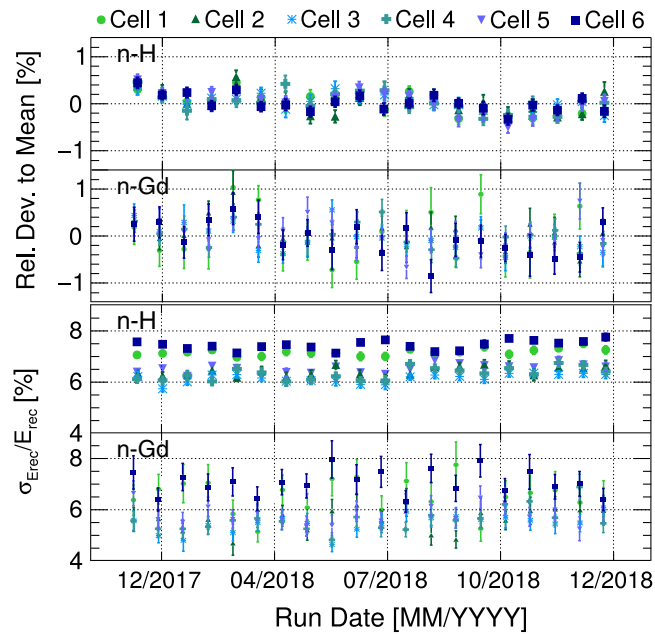

FIG. 13. Top (bottom): Time evolution of the peak position (sigma) of a Crystal Ball fit to the energy of captures of muon-induced neutrons by hydrogen and gadolinium for phase-II. Values are given relative to the means (reconstructed energies) of individual cells over time.

### E. Information from continuous energy spectra

The residuals in Figs. 10 and 12 show no other significant correlated systematic uncertainty to be considered. Instead, the cell-to-cell fluctuations are studied to determine the uncorrelated systematic uncertainty.

The simplest estimate would consist in setting a 1% uncorrelated uncertainty on the calibration coefficients (cf. grey area in Fig. 10). However, for a more robust analysis two steps beyond this approach were taken. First, higher order terms were allowed to fit the energy dependence of the residuals, and second, data complementary to the calibration sources were added in a combined fit of the energy scale in each cell. In principle, the absolute position of the 2.2 MeV peak from the data of cosmic n-H captures would be a good candidate to complement the sources, but testing the energy scale at the (sub-)% level requires a quite accurate control of the simulated distribution of vertices, which is still not demonstrated for these complex events induced by muon spallation events in structures outside the TG volume. At the current stage of the simulation, we estimate that the absolute position of the n-H capture peak induced by cosmic rays is affected by systematic uncertainties comparable to the small distortions we want to test.

Therefore, the analysis of another complementary source of calibration data has been developed: the $\beta$-decay spectrum of the $^{12}$B isotope. This unstable nucleus is naturally produced by the exposure of the liquid scintillator to cosmic rays. Two dominant processes are involved with the $^{12}$C atoms of the liquid as common target: spallation reactions and capture of negative muons stopped in the TG volume. The produced $^{12}$B nuclei undergo $\beta$-decay with a half-life of 20.20 ms and a $Q$-value of 13.37 MeV, thus largely covering our energy range of interest.

The combination of a high muon rate in STEREO and the long lifetime of $^{12}$B makes it challenging to select time-correlated pairs of events. To optimize the selection, the discrimination power of the muon capture process is exploited. Muon candidates are chosen in the [45, 120] MeV energy range corresponding to a distribution of stopping points covering a fairly large fraction of the cell volume, but rejecting traversing muons and cosmic showers. The $\mu^-$ capture process is further enhanced by rejecting signals from decaying muons identified by their associated Michel-electron. Their characteristic signature is defined as a coincidence between a muon candidate followed by an event in the [5, 70] MeV energy range within 6 μs. Then, the boron candidate is searched in the same TG cell and in a [2, 35] ms time window after the muon candidate. The spallation neutrons are an important source of fake $^{12}B$ events through their capture in the detector. This contribution is mitigated by requiring no muon event in 200 μs before the $^{12}B$ candidate. The remaining background is subtracted by looking for accidental coincidences in ten off-time windows (cf. Sec. IX A). That way, $793.2 \pm 3.5$ boron candidates per day in 573.8 days of analyzed data. The signal-to-background ratio is between 0.1 and 0.8 in the [3.75, 14.50] MeV range. This selection of the muon capture process also cures the main difficulty of the simulation of n-H events which is the accurate control of the vertex distribution. Here, the location of stopping muons relies on the simpler physics of muon energy losses. Muon angular and energy distributions are well known at the surface level and have been cross-checked by on-site measurements [25].

As for the radioactive sources, the quality of the energy scale is determined by comparing this experimental spectrum to the simulation. In the $^{12}B$ MC event generator, the $\beta$-transitions to the ground state of $^{12}C$ (98.22% branching ratio) and to the first two excited states have been included with their associated $\gamma$ and $\alpha$ particle emissions, accounting for a total branching ratio of 99.94%. These three transitions are of allowed type with well-known endpoints. Thus, the theoretical uncertainties on their spectrum shape are dominated by the following correction terms to the Fermi theory: weak magnetism and radiative corrections. For both terms, the $1\sigma$ uncertainty band is obtained by comparing simulated spectra with different scenarios of these corrections. The expression of the weak magnetism correction is varied between the original approximate formula in [62] and a more recent calculation [63]. In a recent publication of Daya Bay [64], it is argued that radiative corrections should not be applied to the simulated $^{12}B$ spectrum because the virtual loop corrections do not change the detected kinematics and the energy taken by the emitted real photons is actually deposited in the liquid scintillator. In the case of the smaller STEREO detector, real photons could escape the active volume and their $1/E$ probability density also induces some quenching. Therefore, the variations of the spectrum simulated with or without the radiative corrections were taken as a $\pm 1\sigma$ envelope around a mean spectrum, defining the reference. Another source of uncertainty is the distribution of muon stopping points given as input to the event generator. The nominal $X$, $Y$, and $Z$-distributions were taken from the muon simulation of the STEREO experiment, applying the same energy and topological cuts as in the data analysis. Mild dependences on $X$ and $Y$ are obtained, while a clear correlation between $Z$ and the selected prompt energy shows up, as expected. The configurations with all vertices at the center of the cells and with uniform vertices in the whole volume were also simulated to test the sensitivity of the predicted spectrum shape. As these are extreme cases, the associated uncertainty band was taken as a $\pm 2\sigma$ level. All the above systematic effects are well fitted by linear distortions of the simulated $^{12}B$ spectra and the uncertainties are propagated using a covariance matrix formalism. The uncertainties associated with the radiative corrections and the vertex distribution are dominant and comparable in magnitude.

Finally, a possible contribution of $^{12}N$ $\beta$-decays was investigated. This isotope can be produced from $^{12}C$ nuclei either by charged-current interaction of a $\mu^+$ on $^{12}C$ or charge exchange after proton production by a spallation process (although this last process should be suppressed by our selection cuts). The half-life of $^{12}N$ is 11 ms with an endpoint of 17.38 MeV, thus passing the boron selection cuts. The higher endpoint provides some sensitivity to this contribution and the experimental spectrum is fitted with a weighted sum of the two $\beta$-emitters. The fitted contribution of $^{12}N$ is found to be negligible as it amounts to $(1.2 \pm 0.3)\%$ of the integral of the total spectrum. A reasonable agreement between the measured and simulated shapes is obtained (cf. Fig. 14), with a $\chi^2/\text{ndf} = 40/25$ corresponding to a p-value of 0.03.

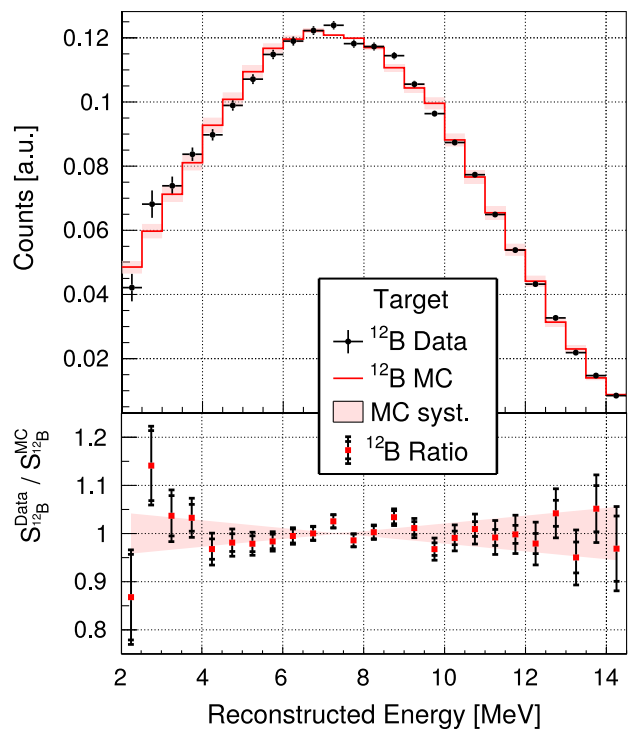

FIG. 14. Top: comparison of the experimental (black circles) and the simulated (red solid line) spectral shape of the selected boron candidates in the TG. The uncertainties of the data are statistical and the red shaded area shows the systematic uncertainty of the simulation. Bottom: data/MC ratio. The uncertainty bars on each ratio point show the statistics-only and total uncertainties.

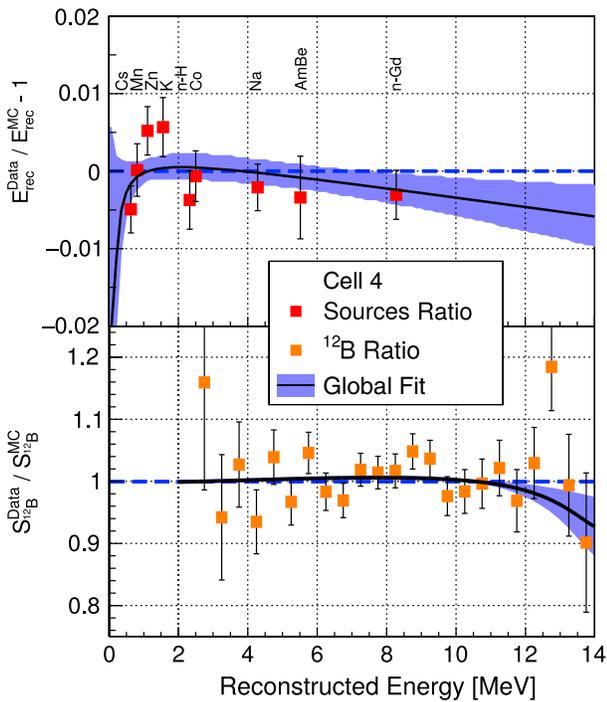

FIG. 15.   Example of a combined fit of sources (top) and boron data (bottom) in cell 4 with a distortion of the energy scale. The experimental energy scale is assumed to be a second order polynomial of the simulated energy scale. The red (orange) points are the Data/MC ratios of sources (boron) data and the blue shaded area is the uncertainty band of the fit.

### F. Combined analysis of all calibration data

Separately in each cell, a combined fit of the boron spectrum (cf. Sec. VI E) and the radioactive source data (cf. Sec. VI C) with a distortion of the energy scale is performed. The residuals of radioactive sources are directly probing the discrepancies between the experimental and simulated energy scales. In the case of the continuous boron spectrum, the impact of a distortion of the energy scale is propagated following the formalism described in Ref. [60]. Polynomial functions of different degrees (between 1 and 4) are fitted to the combined data to probe possible energy nonlinearity models. A typical example of a combined fit by a second-order polynomial is illustrated in Fig. 15. The corresponding distortions of the detected IBD spectra are again calculated with the formalism of Ref. [60] and are depicted in Fig. 16. We find the fit results of cells 1 to 5 in better agreement among each other as compared to cell 6. Depending on the cell and the model, the amplitude of the fitted distortion varies, but its shape remains similar for all fits and is always contained within the above-mentioned simple model of $\pm 1\%$ uncertainty on the linear calibration coefficient. Therefore, we take this envelope, mostly determined by the deviation of cell 6 ($-1\%$), as a $1\sigma$ uncertainty with 100% energy-bin-to-energy-bin correlation and no cell-to-cell correlation. Testing the impact of a cell-to-cell uncorrelated bias at the level of $\pm 1\%$ in a pseudoexperiment, we find the results of the oscillation analysis to be stable. This study completes the list of energy-scale-related uncertainties used for the oscillation analysis, reported in Table II.

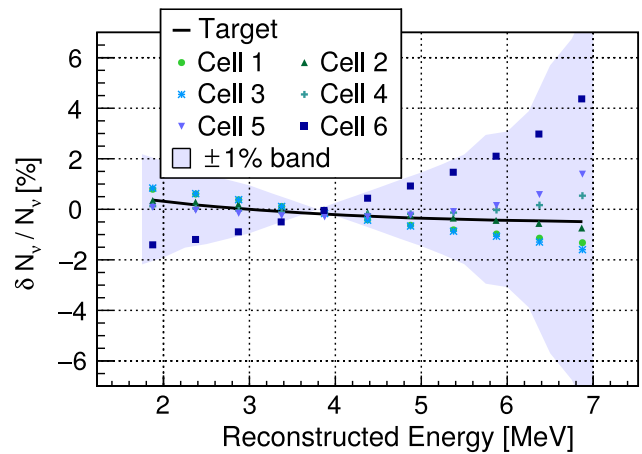

FIG. 16.   Impact of the fitted distortions of energy scale on the detected IBD spectrum for each cell and for the entire TG. The limits of the blue shaded area are drawn from a $\pm 1\%$ bias on the linear calibration coefficients.

## VII. EVENT SELECTION

The first step of background discrimination is done by applying cuts. Each cut aims for an optimal compromise between the signal acceptance and background rejection by minimizing the relative statistical uncertainty on the IBD rate. At the same time, cuts are chosen so that cut results are stable under small cut variations. Using an MC simulation of reactor antineutrinos and taking solid angle effects into account, the spectral distortion caused by a varying cut acceptance with energy is evaluated for each cut and each cell. Distortions, which are too strong, by energy or cell, may fake or obscure patterns of neutrino oscillations. In the oscillation analysis (cf. Sec. XI), the cut acceptances are included by applying the same selection cuts to data and MC events.

All cuts are given in Table III. In the following, we discuss each cut individually. We report cut acceptances as

$$a_{\text{cut}} = \frac{N_{\text{all cuts}}}{N_{\text{all cuts w/o studied cut}}}. \quad (3)$$

TABLE III.    Selection cuts for IBD candidates and their acceptance $a_{cut}$ [cf. Eq. (3)].

| Type | # | Requirement for passing cut | $a_{cut}/\%$ |
|---|---|---|---|
| Energy | 1 | $1.625\text{ MeV} < E_{prompt}^{detector} < 8.125\text{ MeV}$ | 89.2 |
| | 2 | $4.5\text{ MeV} < E_{delayed}^{detector} < 10.0\text{ MeV}$ | 75.9 |
| Coincidence | 3 | $2\text{ μs} < \Delta T_{prompt-delayed} < 70\text{ μs}$ | 95.5 |
| | 4 | $\Delta X_{prompt-delayed} < 600\text{ mm}$ | 99.3 |
| Topology | 5 | $E_{prompt}^{cell} < \begin{cases} 1.0\text{ MeV, neighbor cell} \\ 0.4\text{ MeV, other cell} \end{cases}$ | 98.6 |
| | 6 | | 99.6 |
| | 7 | $E_{delayed}^{Target} > 1.0\text{ MeV}$ | 97.9 |
| Rejection of muon-induced background | 8 | $\Delta T_{muon-prompt}^{veto} > 100\text{ μs}$ | |
| | 9 | $\Delta T_{muon-prompt}^{detector} > 200\text{ μs}$ | |
| | 10 | $\Delta T_{before\ prompt} > 100\text{ μs}$ and $\Delta T_{after\ delayed} > 100\text{ μs}$ for all events with $E_{event}^{detector} > 1.5\text{ MeV}$ | |
| | 11 | $\dfrac{Q_{PMT\ max,\ prompt}}{Q_{e,prompt}} < 0.5$ | 99.3 |

We define the vertex cell of an event as the TG or GC cell which exhibits the largest reconstructed prompt energy.

Cut #1 selects prompt IBD candidates based on the expected energy spectrum of reactor antineutrinos. We find an acceptance of 89.5% of the lower cut at 1.625 MeV and 99.7% of the upper cut at 8.125 MeV. Both cuts show the same acceptance for central cells, but less events are accepted in the outer cells (up to ∼0.2% for cell 1). The lower cut is chosen to avoid background contributions, mainly from neutron-activated $^{41}$Ar, which is present during reactor-on periods only and is fluctuating in time. In the final extraction of IBD events described in Sec. X, only the energy bins up to 7.125 MeV are considered due to the low event statistics and low signal-to-background ratio at higher energy. A cut at this energy has an acceptance of 98.5%.

Cut #2 selects delayed neutron captures by Gd, releasing $\gamma$-rays with a total energy of ∼8 MeV. With this technique, the number of accidental coincidences, which are mainly caused by low energy events, is strongly reduced. The lower cut value of 4.5 MeV is optimized to include a wide part of the tail of the $\gamma$-cascade of the Gd events while avoiding energy depositions of neutron captures by hydrogen at ∼2.2 MeV. Hydrogen is mainly present in the detector as part of the organic scintillator. The lower cut has a signal acceptance of 76.0% with a small difference for cells 1 and 6 of ∼0.6%. The rejected IBD events are mostly those where the neutron is captured by hydrogen. The upper cut value at 10 MeV is fixed to the end of the delayed spectrum in detected energy (see discussion in Sec. VI B). It has an acceptance of 99.9%. While the upper cut shows a flat dependence on energy and cell number, the lower cut preferably removes events at low prompt energy leading to a 1.4% evolution over the prompt energy cut range for all cells. Looking at the vertex distribution of rejected events, one finds that mainly events close to the bottom and top of the TG are rejected. The corresponding delayed events are likely to deposit some energy in inactive detector components, since the GC does not extend above and below the TG (cf. Fig. 2). Since the positrons have a similar spatial distribution, they more likely deposit a part of their energy in the detector compared to events in the detector bulk, due to leaking annihilation $\gamma$-ray or positron trajectories. Thus, not applying this cut adds events where a part of the prompt energy has been lost [cf. Eq. (3)]. Therefore, the formal acceptance of the cut is lower at low prompt energy than at high prompt energy.

Cut #3 allows the rejection of background by exploiting the delay of the neutron capture in the TG due to thermalization and the time constant of the capture process. For the STEREO TG scintillator this time constant is found to be 16 μs. The lower threshold at 2 μs allows the rejection of decays of stopping muons and has an acceptance of 98.0%. The decay of stopping muons can mimic an IBD event when the energy of the end of the ionization track of the muon falls into the prompt energy window and the energy of the produced Michel-electron falls into the delayed energy window. As the lifetime of muons is rather short at 2.2 μs, this background can be reduced by the lower time cut. The upper time cut is fixed to 70 μs where the contribution of accidental coincidences becomes dominant over physically correlated events and has an acceptance of 97.6% for cells 2 to 5, 97.4% for cell 1 and 96.7% for cell 6. While the lower cut shows no effect on energy or cell number, the upper cut shows a 0.7% effect which is slightly bigger (1.0%) for cell 6. The different value for cell 6 can be explained by considering the directionality of incident antineutrinos that is propagated into the IBD neutrons. Cell 6 is the last cell before the GC (cf. Fig. 1) and, therefore, there is a higher chance that neutrons have a trajectory reaching the GC. Since the capture time constant is longer in the unloaded GC scintillator, it can take more time before the neutrons reenter the TG and get captured.

Cut #4 exploits the fact that accidental coincidences are not correlated in space. Thus, they show, on average, a larger distance between prompt and delayed vertices as compared to an IBD event, where positron and neutron originate from the same interaction vertex. Consequently, the requirement of a maximal distance between vertices along the $X$-axis (cf. Fig. 2) of 600 mm, calculated from the barycenters of the two vertices, rejects predominantly accidental correlations. The barycenters are determined by weighting the positions of all PMTs $j$ within a cell $l$ by their charges $Q_j$, and within the entire detector by the reconstructed energy $E_l$ in their cell $l$. This procedure reduces the effect of light cross-talk. The cut has an acceptance of 99.3% and shows negligible spectral distortions.

For cut #5, we require the vertex cell to be a TG cell. Then, by limiting the energy in each cell neighboring the vertex cell to 1.0 MeV, and each cell not neighboring the vertex cell to 0.4 MeV, cuts #5 and #6 allow only one of the 511 keV $\gamma$-rays of an IBD-positron annihilation to be detected in a given neighbor cell. Two escaping 511 keV $\gamma$-rays are unlikely to be detected in the same neighbor volume, and the margin between 511 keV and the 1 MeV cut is safe against the light cross-talk from the vertex cell to its neighbors. The acceptances of the cuts #5 and #6 are 98.6% and 99.6%, respectively. Cells 1 and 6 show a 0.3% higher acceptance, mainly because other cuts already removed events at the border to the GC and those cells are mainly surrounded by GC cells. The energy distortion of these cuts is 2.9% and 0.4%, respectively, over the full energy range. It is caused, e.g., by fluctuations in the energy transferred to the neighbor cell due to energy cross-talks being larger at higher prompt energies.

By requiring the deposited energy in the entire Target to be larger than 1.0 MeV for the delayed event, cut #7 allows to reject high energy background events coming from the sides into the detector. Those contribute mainly to accidental coincidences. The cut accepts 98.8% of IBD events for cells 2 to 5, but shows a lower acceptance (96.0%) for cells 1 and 6 since one of their neighbor cells is a short GC cell instead of a TG cell (cf. Fig. 2). The energy distortion is at the 0.3%-level for central cells and 1.0% for cells 1 and 6 for similar reasons as for cut #2.

Cuts #8 and #9 introduce a veto-time after a muon candidate is detected by its deposited energy in either the water Cherenkov veto (100 μs) or the detector (200 μs). A muon candidate in the detector is an event with a total energy above 20 MeV. Cuts #8 and #9 reject a large part of muon-induced background such as fast neutrons which can mimic an IBD event by a prompt proton recoil and a subsequent neutron capture of the same neutron or another neutron induced by the same muon. Those events are close in time to the incident muon. Likewise, cut #10 allows to reject correlated background events originating from a multiple neutron cascade with an undetected primary particle by requiring no event with more than 1.5 MeV energy to happen less than 100 μs before or after an IBD candidate. Those neutrons are expected to be close in time. The acceptance of these three cuts is directly corrected via the veto time.

By limiting the ratio between the maximal charge collected by a single PMT and the total charge collected by all PMTs of the vertex cell during a prompt event to less than 0.5, cut #11 allows to reject stopping muons in addition to the lower time cut introduced by cut #3. To be not already rejected as muon event by cuts #8 or #9, a stopping muon needs to deposit only a small amount of energy in the detector, i.e., it can only have a short track in the active detector volume. Consequently, it must have a vertex distribution very peaked at the top of the detector and hence the distribution of scintillation light is asymmetric compared to events in the bulk of the detector. By looking at the distribution of charge between the four PMTs of one cell, stopping muons can be discriminated as they are likely to produce most scintillation light in the vicinity of just one PMT. This leads to a higher ratio between the maximal charge of one PMT and the total charge of all PMTs than for bulk events. This cut has an acceptance of 99.3% and shows a small energy distortion of 0.4%.

Before any selection, we find the ratio of antineutrino interactions in MC to be $39 : 47 : 14$ for TG, GC and acrylic detector components (mainly Buffers), respectively. The ratio is better than $99 : 0 : 1$ after applying selection cuts. The overall acceptance in the TG after applying all cuts is 60% showing an ∼4% relative variation across the prompt energy range. Details can be seen in Fig. 17. The central cells (2–5) show very a similar behavior with an average of 63%, while cell 1 exhibits a few percent lower acceptance due to the loss of neutron detection efficiency at the edge of the detector (cf. Sec. VIII). Cell 6 is affected by the same effect, but has some additional neutron loss due to the directionality of incident antineutrinos that is propagated into the IBD neutrons. The resulting small energy distortion has a common shape for all cells.

In this study, the full set of cuts was compared with the set of cuts reduced by the cut of interest [cf. Eq. (3)] to get rid of the existing correlation between cuts. The systematic uncertainties of the resulting acceptances induced by the cuts #3 to #7 cannot be derived directly from the uncertainties of the used observables because of the large correlations of the cuts. For example, cuts #5 and #6 use the same cell energy reconstruction. Therefore, the systematic uncertainties were estimated for each cell and energy bin by an MC study. For this study, observables used for the selection of these cuts were randomly varied within their estimated uncertainties before being applied to the IBD MC, resulting in a new number of selected IBD candidates. By redoing the procedure multiple times, distributions of selected IBD events in each cell and each energy bin are produced. The resulting means of the IBD distributions are compared to the number of IBDs obtained without any variation and the differences give the resulting systematic uncertainty for each cell and energy bin. For the sum of all cells, the systematic uncertainties are of the order of 0.5% below 4.375 MeV and increase up to 2.0% at 7.875 MeV. As these uncertainties are cell-to-cell correlated, they do not enter the oscillation analysis, which is insensitive to common shifts among detector cells. However, they lay the basis for absolute measurements, to be discussed in future publications. Note that the systematic uncertainties do neither take into account energy scale systematics (cut #1 and #2) nor energy nonlinearities which are taken into account separately in the oscillation analysis (Sec. XI). Moreover, they do not include the delayed neutron detection efficiency (Sec. VIII) which is also taken into account separately in the oscillation analysis.

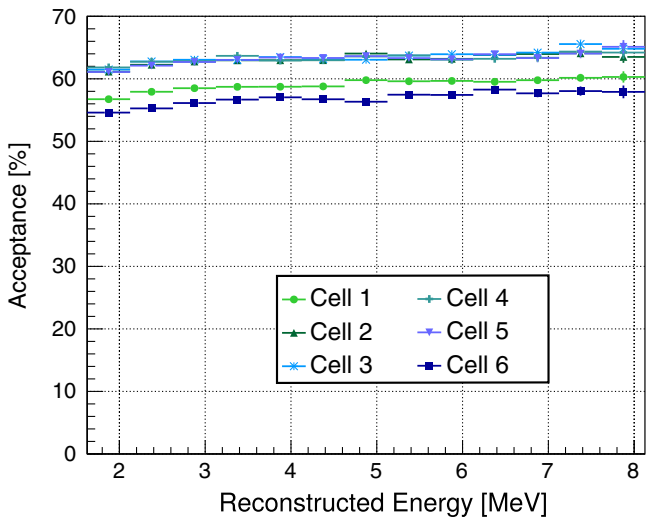

FIG. 17. Spectral distortion in reconstructed prompt energy due to the acceptance of IBD events after applying all selection cuts.

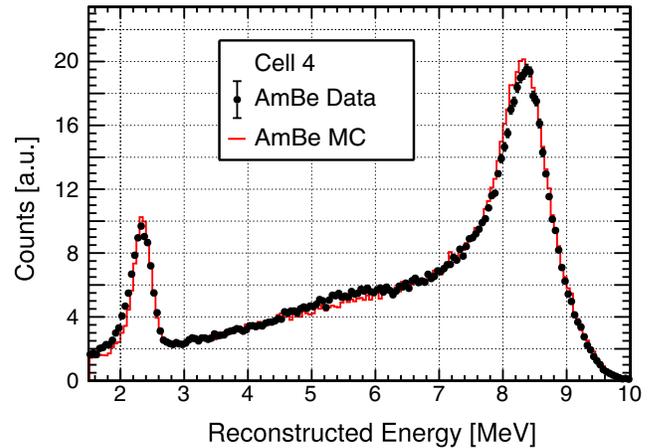

FIG. 18. Spectrum of neutron capture events of a delayed coincidence search for an AmBe source deployed in cell 4. The shown spectrum is obtained as a sum of 5 spectra measured at 5 vertical heights along the Z-axis (cf. Table I). Accidental background has been subtracted.

## VIII. DELAYED NEUTRON DETECTION EFFICIENCY

As it was introduced in Sec. II B, antineutrinos are detected in STEREO by their interaction on a proton, followed by a positron annihilation and a delayed neutron capture. To suppress low energy background, mainly neutron captures by Gd are accepted (cut #2 in Sec. VII). In order to have a good understanding of the antineutrino interaction rate, it is necessary to know the efficiency of the selection of neutron captures by Gd nuclei. In this section, the estimation of this term and its implementation in the simulation framework of STEREO are explained.

The fraction of neutron captures by Gd in the detector is evaluated using a neutron source. To build a neutron detection efficiency map of the TG volume, the source is deployed through the internal calibration tubes into the TG cells at five different heights along the Z-axis of the detector (cf. Table I). In the STEREO experiment, an americium-beryllium ($^{241}$Am/$^9$Be) source is used as a fast neutron emitter. These neutrons are produced in $^9$Be($\alpha, n$)$^{12}$C reactions, where the $\alpha$ particle is created in radioactive decays of $^{241}$Am. In about 60% of the cases, the carbon nucleus is produced in an excited state, and emits a 4.4 MeV $\gamma$-ray in addition to the neutron [65]. To get a background-free neutron capture sample, this prompt high energy $\gamma$-ray is used to perform a delayed coincidence search with the neutron capture. The coincidence selection cuts are applied in a similar way as for the IBD: delayed signals are searched in a time window of 100 $\mu$s after a 4.4 MeV prompt candidate. Also, contributions from random coincidences (accidental background) are selected in a similar way as for IBD events (cf. Sec. IX A). They are subtracted statistically. Figure 18 displays a neutron capture spectrum for an AmBe source, obtained after coincidence search and accidental subtraction. A good agreement between data samples and MC simulations have been reached due to the implementation of the FIFRELIN code [57], improving the description of the de-excitation cascades of the relevant Gd isotopes [58]. The distribution of coincidence times of neutron captures are represented for both data and MC simulations in Fig. 19, showing the agreement of the capture times between both samples.

The efficiency of the detection of neutron captures by Gd is estimated taking into account two different terms. The first is the Gd-fraction ($\varepsilon_{Gd}$), which takes into account the fraction of events captured by Gd with respect to other elements in the detector. This term depends on the abundance of these elements in the liquid scintillator and their corresponding neutron capture cross-sections. In the TG volume, captures by Gd and H dominate this competitive process. Neutron captures by Gd release a $\gamma$-cascade with a total energy around 8 MeV, whereas captures by H are followed by a single $\gamma$-rays of 2.2 MeV. Figure 18 shows that the Gd-capture events ($N_{Gd}$) are mainly distributed with energies between 3 and 10 MeV (Gd-capture peak and its tail). H-captures ($N_H$) are distributed with energies in the range [1.5, 3.0] MeV, where the lower energy cut of 1.5 MeV is established to exclude the region with background [66]. Considering these energy cut ranges, the Gd-fraction is defined as

$$\varepsilon_{Gd} = \frac{N_{Gd}}{N_{Gd} + N_H}. \qquad (4)$$

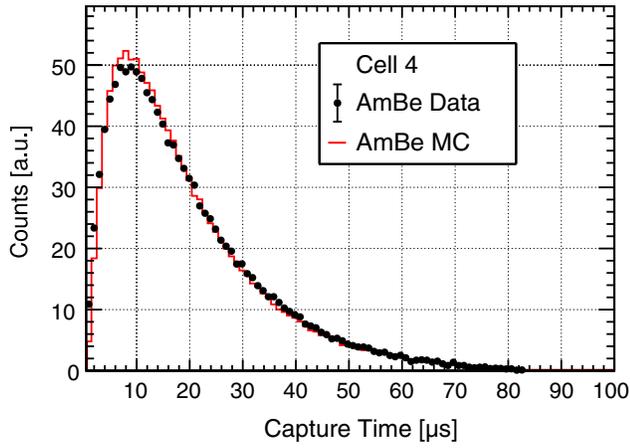

FIG. 19. Distribution of neutron captures over times for an AmBe source deployed in cell 4. The shown distribution is obtained as a sum of 5 measurements at 5 vertical heights along the Z-axis (cf. Table I). Accidental background has been subtracted.

However, this neutron capture fraction does not take into account all the selection cuts applied by the coincidence search of an IBD event (cf. Table III). The main selection cuts affecting the delayed events are the energy cut #2, the topological cut #7, and the coincidence time cut on the delayed event #3. In this way, the IBD selection cut efficiency

$$\varepsilon_{IBD} = \frac{N_{cut\#2 \,\cap\, cut\#3 \,\cap\, cut\#7}}{N_{Gd}} \qquad (5)$$

is defined as the fraction of the Gd-capture events fulfilling all of these cuts. This yields the total delayed detection efficiency

$$\varepsilon_{tot} = \varepsilon_{Gd} \cdot \varepsilon_{IBD}. \qquad (6)$$

Neutron efficiency is evaluated in the same way for AmBe calibration data samples and MC simulations at the same source positions. The efficiency terms have been estimated in both samples, and compared to evaluate possible discrepancies caused by imperfections of the MC simulation.

The ratio between data and MC efficiency terms are named correction coefficients and are defined as

$$c_x = \frac{\varepsilon_x^{data}}{\varepsilon_x^{MC}} \qquad (7)$$

where $x \in \{Gd, IBD\}$ denotes the specific efficiency term.

For the IBD selection cut efficiency, an averaged value of $c_{IBD} = 0.9985 \pm 0.0059$ has been obtained from all the calibration points inside the detector. The uncertainty includes systematic effects from time variations of the 25 detector calibration points in time, inhomogeneities, and displacements of the source position within the deployment accuracy. The averaged value is consistent with unity within one standard deviation.

The Gd-fraction correction coefficient $c_{Gd}$ is not constant throughout the TG volume mostly caused by an imperfect model of the mobility of thermal neutrons in the simulation. It decreases toward the edges of the TG, as can be seen in Fig. 20. Values differ by 2% between the central point of the detector and the border regions. From these calibration points, a two-dimensional fit function is derived. The two lateral directions toward the short GC cells can be described by a Subbotin distribution (generalized normal distribution [67])

$$C_{Gd}(x) = \exp\left[-\left(\frac{|x-\mu|}{\sigma}\right)^\beta\right] \qquad (8)$$

allowing for a plateau toward the center of the detector. In Eq. (8), $\mu$ denotes the mean of the distribution (set equal to zero in the fit), while the parameters $\sigma$ and $\beta$ allow scaling and shaping the distribution, respectively. The distribution is already given in a renormalized form, which is used in the fit. The fitted distribution is represented in Fig. 20 with a grey line. A similar effect is expected toward the long GC cells. Therefore, the obtained fit function from the X-axis is also applied to the Y-direction of the detector using an adapted plateau length. With this two-dimensional function, correction coefficients $c_{Gd}$ have been obtained per cell. Since the STEREO detector is not aligned with the antineutrino flux from the reactor core (cf. Fig. 1), its orientation has to be taken into account in the computation of the correction coefficients. Weighting the two-dimensional function by the expected distribution of IBD vertices, the $c_{Gd}$ coefficients are found to be 0.9650 and 0.9846 for border and central cells, respectively. These results are slightly lower than results obtained with the AmBe calibration points since the border effect along the Y-axis is now taken into account.

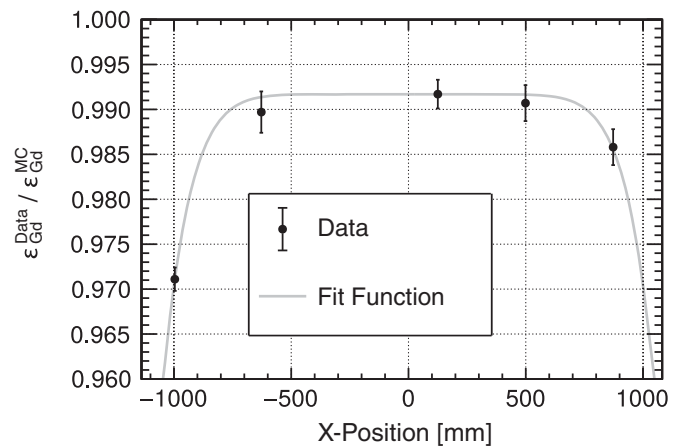

FIG. 20. Ratio between the Gd efficiency of data and MC events $c_{Gd}$ [cf. Eq. (7)] for an AmBe source averaged over 5 vertical heights in each TG cell (cf. Table I). The grey line represents the fit function as detailed in Eq. (8).

TABLE IV. Correction coefficients $c_{tot}$ for the delayed neutron efficiency. Two different uncertainties (cell-to-cell uncorrelated and correlated) are computed taking into account systematic effects of the $c_{Gd}$ and $c_{IBD}$ contributions [66].

| Cell | $c_{tot}$ | $\delta_{uncorr}(c_{tot})$ | $\delta_{corr}(c_{tot})$ |
|---|---|---|---|
| 1 | 0.9635 | | 0.0041 |
| 2 | 0.9828 | | 0.0015 |
| 3 | 0.9831 | 0.0084 | 0.0013 |
| 4 | 0.9831 | | 0.0013 |
| 5 | 0.9829 | | 0.0015 |
| 6 | 0.9643 | | 0.0040 |

Two different types of uncertainties are applied to the $c_{Gd}$ values. On the one hand, a cell-to-cell uncorrelated uncertainty of 0.57% has been obtained considering time fluctuations, source position bias, and residual discrepancies between model and measured values. On the other hand, a cell-to-cell correlated term has been estimated testing a different fit function for the AmBe efficiency ratios. More sensitive to the uncertainty of the slope of the fit function, cells 1 and 6 have the highest correlated uncertainty term (0.41%), whereas the impact on central cells within the plateau region is limited to 0.13% on average (cf. Fig. 20).

The total correction coefficients

$$c_{tot} = c_{Gd} \cdot c_{IBD} \qquad (9)$$

are summarized in Table IV for each cell. Uncertainties are computed considering the previously mentioned systematic effects and cell-to-cell correlations. These correction coefficients contain both, the neutron capture fraction and the impact of the IBD selection cuts. They are taken into account in the oscillation analysis as factors, which correct the normalization of each individual cell (cf. Table II).

## IX. BACKGROUNDS

### A. Accidental background

Uncorrelated, i.e., random, combinations of events can pass the selection cuts described in Sec. VII. Their occurrence depends directly on the frequencies of single prompt and delayed events. The rate and spectrum of these accidental coincidences is evaluated statistically by an off-time method. Each valid prompt candidate is shifted in time by $t_1 = 2$ ms and the IBD selection is again applied. The time shift is chosen to be large compared to the neutron capture times in the liquids, such that any valid delayed event following the shifted prompt is not correlated to it. The formed pair is then registered in the accidental sample. This step is repeated with $g_{max} = 10$ different times $t_g = g \cdot t_1$, $1 < g \leq g_{max}$ in order to increase the statistical precision of the accidental background contribution. This method is applied along with the IBD search and monitors then accurately—for each cell and each energy bin—any change of this accidental background, caused by, e.g., changes in the configuration of the neighboring instruments. Variations in the accidental rate up to a factor 4 were recorded between reactor-on and reactor-off periods. We find the prompt spectrum of accidental coincidences to be dominated by low energy γ-ray events, and reflecting the single events spectrum (cf. Fig. 21). Contributions above the highest γ-energy peak of natural radioactivity at 2.6 MeV from Tl mainly come from the cascades following neutron captures on Gd and high-energy γ-rays from neutron capture on surrounding materials (Al, Fe) during reactor-on periods, due to neighboring instruments. Correlated coincidences by, e.g., boron decays (cf. Sec. VI E) impose a negligible background.

The accuracy of the off-time method was validated by performing an IBD event selection with a long delay time of 2 ms (cut #3, Sec. VII). Looking at the spectrum of delay times above 1 ms, the study shows that the estimation of the accidental background rate exhibits subpercent bias. Any potential residual bias propagates to the signal extraction, and is maximal for high accidental rates and large differences between reactor-on and reactor-off. In the most unfavorable conditions—encountered in the lowest energy bins—this systematic on the rate of neutrino events is estimated to be less than 0.5%, and is negligible in the higher energy bins.

### B. Model of correlated background

Correlated background refers to any physically correlated process similar to an IBD signature. While the accidental background can be estimated along the data flow irrespective of the reactor status, this correlated background contribution is characterized from the periods where the reactor is turned off. It is therefore mandatory that the combination of the shielding of the detector and the selection cuts suppresses all correlated background connected to the reactor activity. A stringent upper limit of this kind of background is presented in Sec. IX C. We discuss here the background induced by the cosmic rays, virtually the single source of correlated background.

The residual correlated background after the IBD selection can be further classified as electronic recoils (weakly ionizing particles, expected for the IBD signal) and proton recoil background (strongly ionizing particles). The latter is the dominant one and arises from fast neutrons interacting in the liquid, inducing a prompt nuclear recoil signal followed by a valid delayed signal from their capture.

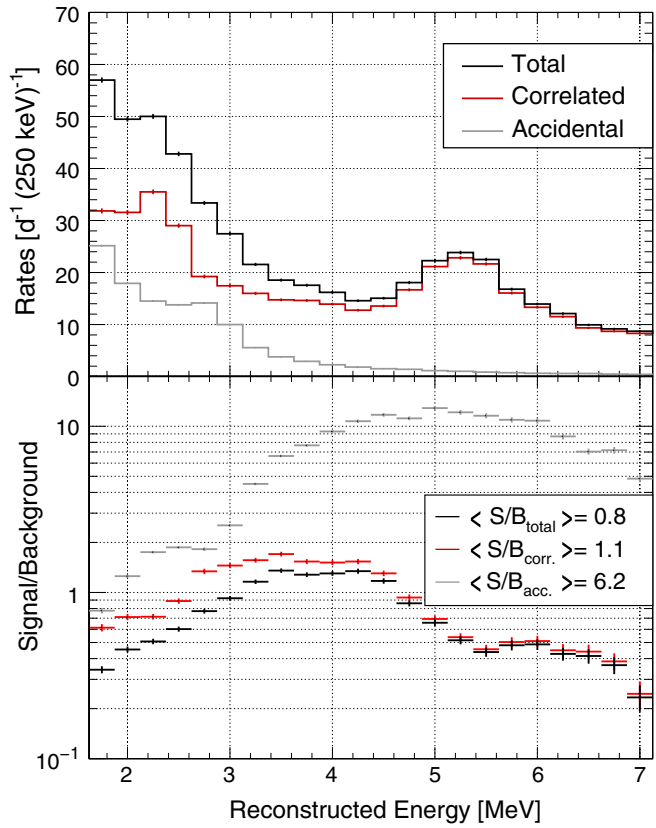

FIG. 21. Top: background spectrum of correlated and accidental (during reactor-on periods) contributions, defined as the rate integral in the electronic recoil component ($Q_{tail}/Q_{tot} < \mu_\gamma + 2\sigma_\gamma$, cf. Fig. 22). Bottom: measured signal-to-background ratio with respect to the background spectra shown above. The average values given in the legend are calculated from the data points weighted by the signal.

Their rejection is based on the pulse shape discrimination (PSD) capability of the liquid, whose typical distribution for reactor-off correlated IBD candidates of one energy bin and one cell is shown in Fig. 22. The population at low $Q_{tail}/Q_{tot}$ is made up from IBD events, correlated electronic background induced by cosmic rays, and accidental coincidences. The corresponding particles produce electronic recoils in the scintillator, leading predominantly to excitations of $\pi$-electrons of the scintillator in the singlet regime. This leads to fast fluorescence pulses. The second population at larger $Q_{tail}/Q_{tot}$ is dominantly caused by muon-induced fast neutrons. Those neutrons produce proton recoils leading to excitations in the triplet regime. This leads to longer de-excitation times. The delay between the start of the integration time window of the entire scintillation pulse (resulting in $Q_{tot}$) and the start of the integration time window of the tail of the scintillation pulse (resulting in $Q_{tail}$) was chosen such that the best separation of the two populations in terms of $Q_{tail}/Q_{tot}$ was achieved.

Changes in temperature change the liquid density and may thus influence its PSD properties. Indeed, we found a linear anticorrelation between the PSD position and the temperature of the liquid, $\Delta\mu_\gamma/\Delta T \sim -0.1\sigma_\gamma/\text{K}$ [68]. The temperature variation can reach 3°C on a two months scale (typical reactor-on period). To correct from these drifts, the observable $Q_{tail}/Q_{tot}$ is monitored along time for each energy bin and each cell using the high-statistic sample of single events triggering the detector. Such a reference distribution is shown on the bottom plot of Fig. 22. Dominated by electronic recoils, this population is fitted by a normal distribution, where the extracted parameters $\mu_\gamma$ and $\sigma_\gamma$ denote the mean and the standard deviation of the electronic recoil population. The single events thus provide a precise monitoring of the temperature dependence. The correction of a global offset of the PSD distributions is implemented on a daily basis. Electronic recoils are then defined with respect to the singles reference as $Q_{tail}/Q_{tot} < \mu_\gamma + 2\sigma_\gamma$ and proton recoils as $Q_{tail}/Q_{tot} > \mu_\gamma + 2.5\sigma_\gamma$ (cf. Fig. 22). Note that this separation is only used for investigations of the background, but not for the extraction of the neutrino signal presented in Sec. X.

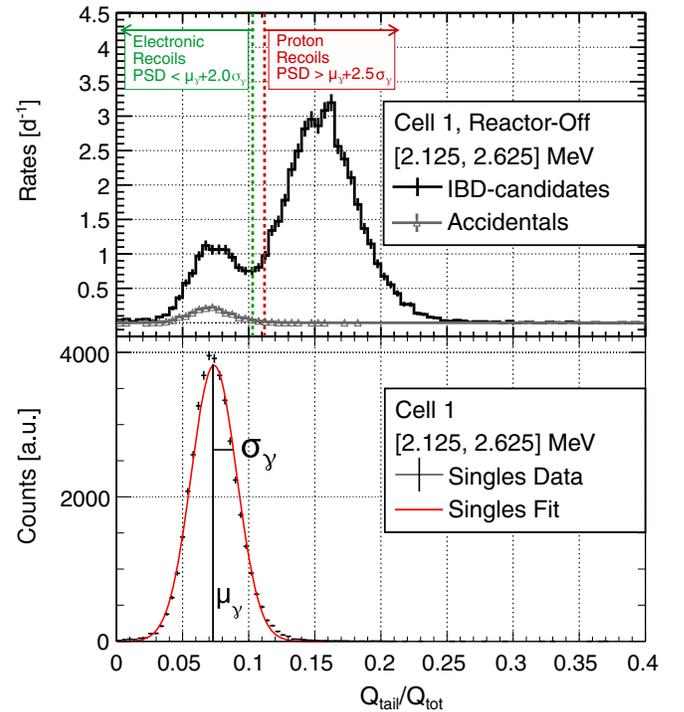

FIG. 22. Top: PSD distribution of the IBD candidates (black) and the accidental contribution (grey) obtained during the reactor-off periods of phase-II in the first cell and for prompt energies in [2.125, 2.625] MeV. Bottom: one-day PSD distribution of single events, dominated by electronic recoils. The extracted parameters $\mu_\gamma$ and $\sigma_\gamma$ allow to define a PSD cut to separate electronic recoils from proton recoils.

The prompt background spectrum contaminating the electronic recoil component can be obtained by integrating the rate of events for $Q_{tail}/Q_{tot} < \mu_\gamma + 2\sigma_\gamma$ in each energy bin. The resulting distribution is presented in Fig. 21. Series of neutron captures in the liquid surviving the isolation requirement (cut #10) are responsible for the purely electromagnetic contamination peaked at 2.2 MeV. They are associated to neutron captures by hydrogen. The prominent feature around 5.4 MeV arises from the $^{12}C(n, n'\gamma)^{12}C$ reaction in the liquid. Unlike the previous one, this contribution is not purely electromagnetic because of the presence of a coincident neutron recoil along with the 4.4 MeV $\gamma$-ray in the prompt signal. Combining these background spectra with the shape of the detected IBD spectrum (cf. Sec. X), the signal-to-background ratios for the different contributions are illustrated in Fig. 21.

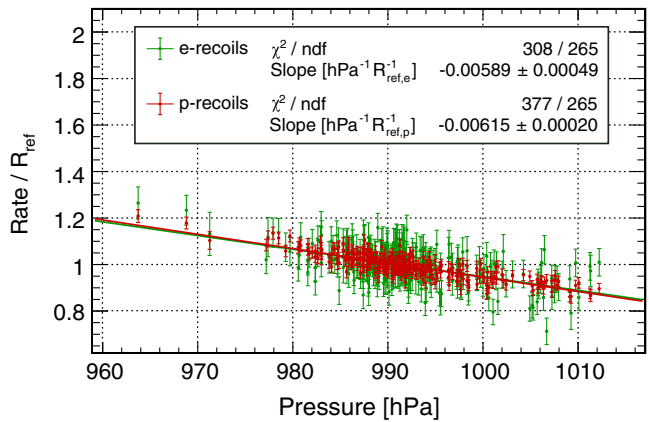

FIG. 23. Relative rates of IBD candidates from reactor-off periods of phase-II as a function of the atmospheric pressure. A compatible relative dependence for electronic (green) and proton (red) recoils is found. The set of electronic and proton recoils in this plot is built from all IBD candidates having a $Q_{tail}/Q_{tot} < \mu_\gamma + 2\sigma_\gamma$ and $Q_{tail}/Q_{tot} > \mu_\gamma + 2.5\sigma_\gamma$, respectively (cf. Fig. 22).

### C. Stability of background model

The correlated background was found to be sensitive to environmental conditions, as expected, because of its cosmic-induced origin. As shown in Fig. 23, rates of electronic recoil and proton recoil events are correlated to the atmospheric pressure. The linear correlation coefficient is found to be compatible between the two populations. A dependence on the filling level of the water pool above the reactor core was also observed, but the greatest effect (due to a 7 m change in the water level) is 16 times weaker than due to a 30 hPa atmospheric pressure variation.

To study the influence of these changes on the PSD distribution, the reactor-off dataset is subdivided into subsets of data, classified by atmospheric pressure and filling level of the water pool above the reactor core. The shape of the PSD distributions remains very stable in all conditions of atmospheric pressure (cf. Fig. 24) and water level (cf. Fig. 25). The fit of a single normalization factor $a$ is enough to accurately overlap the distributions of two subsets of data with opposite variations of the environmental parameters.

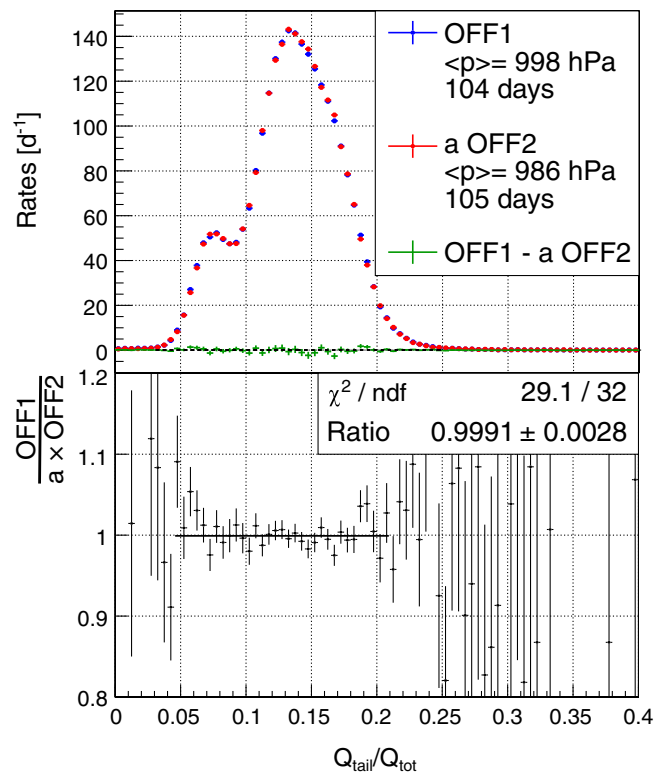

FIG. 24. Top: comparison of pulse shape distributions for the reactor-off dataset of phase-II split into two bins of high and low atmospheric pressure (this figure) and for the two observed values of filling level of the reactor water pool (cf. Fig. 25). The distributions for different atmospheric pressures (this figure) were recorded at the same level of the water pool. The distributions for different levels of the water pool (cf. Fig. 25) have an average pressure difference of 1.6 hPa. In both figures, the spectra are normalized to their acquisition times, while the OFF2 spectra are additionally scaled by parameters $a$. These parameters are determined from fits to the OFF1 spectra allowing only for scaling of the OFF2 spectra. In both figures, the parameters $a$ account for the pressure difference between the datasets. In both fits, they are found to be compatible with the trend presented in Fig. 23. Bottom: ratio between the OFF1 and the scaled OFF2 spectra. In both figures, the spectra are compatible in shape within uncertainties.

Looking at the PSD distribution of all events measured with an AmBe source, we find a small increase of the distance between the peaks of the distributions of electronic and proton recoils in time. For our dataset, this increase was about 1% over the duration of phase-II. The alternation between reactor-on and reactor-off measurements largely reduces the influence of this effect on the extracted antineutrino rates. The remaining effect is found to be cell-to-cell correlated not affecting the oscillation analysis, which is insensitive to common shifts among detector cells.

The best-fit value of the normalization factor $a$ between reactor-on and reactor-off periods (analogous to the comparisons of different reactor-off datasets in Figs. 24 and 25) can serve as a probe of potential reactor-induced background. Such a correlated background could be induced by fast neutrons producing a prompt proton recoil and a delayed capture signal. If sizeable, it would prevent the use of the background PSD spectra measured during reactor-off periods as models of background shape for the reactor-on periods. From the sensitivity to atmospheric pressure shown in Fig. 23 and the equivalent measurement for the water level in the pool, it is possible to project all reactor datasets to the same environmental parameters. Once these corrections are applied, the relative difference of proton recoil rates between reactor-on and reactor-off phases is presented in Fig. 26. Data points at high energy are compatible with zero while points below 3.5 MeV are incompatible with the null hypothesis (more than 3 standard deviations for the first four bins together). In the worst case, the excess exists only for proton recoils and it would directly propagate into an underestimation of the IBD rates in the first energy bins since the antineutrino extraction method assume a constant shape of the PSD distribution (cf. Sec. X). In the best case, the excess is the same for proton and electronic recoils and it would not affect the IBD rates. We consider fast neutrons from the reactor or a neighboring instrument (cf. Fig. 1) as the most likely origin of a hypothetical additional background, producing prompt proton recoils at low energy. Allowing for an offset to account for a possible residual bias in the projection to the same environmental parameters, the excess is fitted by a power law

$$R(E) = u \cdot E^{-v} + w \tag{10}$$

where $E$ is the reconstructed energy and $u$, $v$, and $w$ are free fit parameters. Based on the most likely scenario of a residual reactor-related fast neutron background, we correct for the full bias taking into account the energy-dependent signal-to-background ratios (cf. Fig. 21). The correction is 3.6% and 1.8% in the first two energy bins, respectively, and is negligible in higher bins. We get the same results when fitting the excess with an exponential function. However, if the excess were to be the same for proton and electronic recoils as expected in other scenarios, the IBD rates would not be affected. To cover this case, an uncertainty equal to the size of the correction is conservatively added in each bin as additional uncorrelated systematic uncertainty on the IBD rates. It is small compared to the statistical uncertainty.

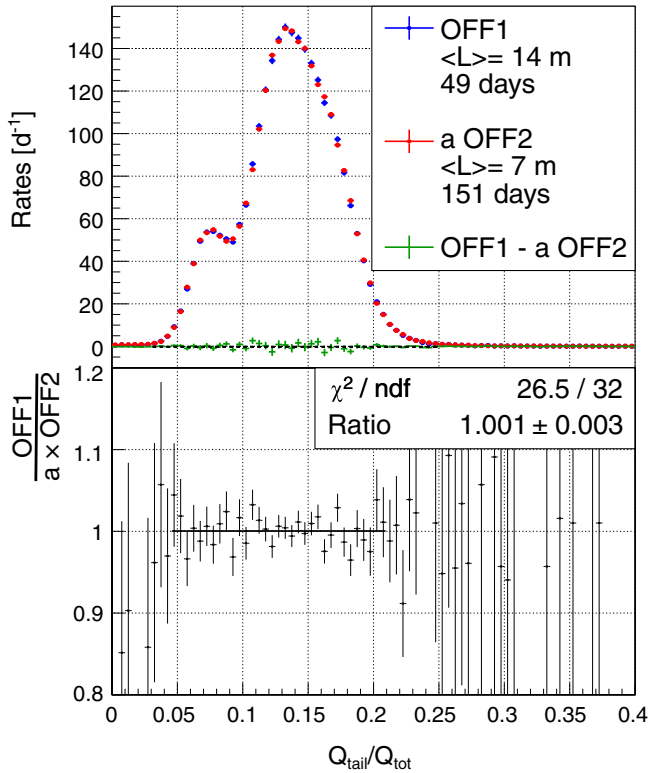

FIG. 25. Comparison of pulse shape distributions of the reactor-off dataset of phase-II for the two observed values of filling level of the reactor water pool. See Fig. 24 for details. The difference in the total number of days between Fig. 24 and this figure arises from the exclusion of 9 days with changing water level in the channel above the detector.

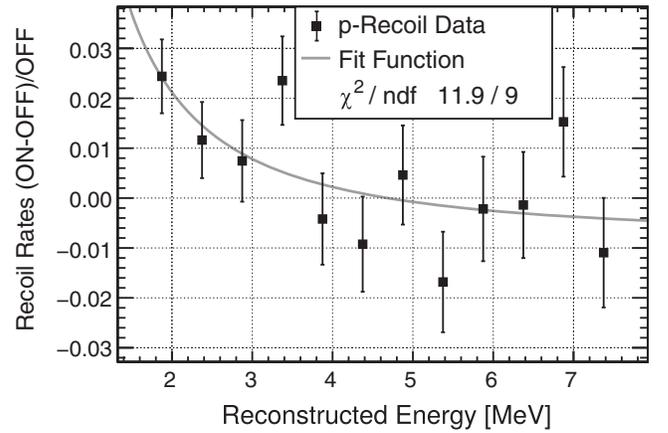

FIG. 26. Relative difference of proton recoil rates, corrected for the measured dependence on the atmospheric pressure and water pool level, between reactor-on and reactor-off periods. To avoid a contribution of IBDs, only events with $Q_{\text{tail}}/Q_{\text{tot}} > \mu_\gamma + 3.0\sigma_\gamma$ are considered (cf. Fig. 22). The grey line represents the fit function as detailed in Eq. (10).

Stable in shape over time, the correlated PSD distributions of IBD candidates obtained from the reactor-off dataset provide a model of the background that can be directly used for the extraction of the antineutrino signal.

## X. ANTINEUTRINO SIGNAL EXTRACTION

Based on the stability of the offset-corrected PSD spectra established in Sec. IX, the rate of IBD candidates is extracted for every cell $l$ and energy bin $i$ from the PSD distributions measured in the reactor-on and reactor-off phases. The PSD distributions of IBD candidates and accidental coincidences measured with reactor-off (reactor-on) are denoted as $OFF_{l,i}$ ($ON_{l,i}$) and $OFF_{l,i}^{acc}$ ($ON_{l,i}^{acc}$), respectively. The PSD distribution of correlated background pairs (without accidental coincidences) is modelled by $m_{l,i}^{cor,OFF}$ for reactor-off periods. For reactor-on periods, the same model is used, only allowing for a normalization factor $a_{l,i}$, which accounts for different mean atmospheric pressures and reactor pool levels (and would also compensate potential imperfections of the dead-time correction).

In contrast, specific models $m_{l,i}^{acc,OFF}$ and $m_{l,i}^{acc,ON}$ are used to describe the accidental coincidences because of the different single spectra. The PSD distribution of IBD events is described by a scaled normal distribution $G^\nu(A_{l,i}, \mu_{l,i}, \sigma_{l,i}^2)$ with mean $\mu_{l,i}$ and standard deviation $\sigma_{l,i}$. The scaling factor $A_{l,i}$ directly measures the IBD rate. Separately for each cell $l$ and energy bin $i$, the four measured spectra are fitted simultaneously for all PSD bins $p$ where $ON_{l,i;p}$ and $OFF_{l,i;p}$ are nonzero (cf. Fig. 27):

$$ON_{l,i;p} = a_{l,i} m_{l,i;p}^{corr,OFF} + f^{acc,ON} m_{l,i;p}^{acc,ON} + G_p^\nu(A_{l,i}, \mu_{l,i}, \sigma_{l,i}^2), \tag{11}$$

$$OFF_{l,i;p} = m_{l,i;p}^{corr,OFF} + f^{acc,OFF} m_{l,i;p}^{acc,OFF}, \tag{12}$$

$$ON_{l,i;p}^{acc} = m_{l,i;p}^{acc,ON}, \tag{13}$$

$$OFF_{l,i;p}^{acc} = m_{l,i;p}^{acc,OFF}, \tag{14}$$

where $f^{acc,ON} = 0.117671$ and $f^{acc,OFF} = 0.117294$ account for the number of time shifts used in the accidentals extraction as well as for the different suppression-times of accidentals in the coincidence window compared to the off-time windows used to measure the accidentals. They are different because accidental signals can be suppressed by correlated events in the coincidence window, but not in the off-time windows. The fit parameters $a_{l,i}$, $A_{l,i}$, $\mu_{l,i}$, $\sigma_{l,i}$, and $m_{l,i;p}^\zeta$, $\zeta \in \{corr, OFF; acc, ON; acc, OFF\}$ are free and are determined independently for the two STEREO phases. Because of the low-statistics regime in most of the PSD bins, a binned log-likelihood maximization is used [68]. The rate of IBDs $A_{l,i}$ is extracted for each cell $l$ and 500 keV energy bin $i$ separately. Compared to the PSD fit used in phase-I [18], the new method does not assume a particular analytical shape of the PSD distribution for background events. The description of the electronic and proton recoil component by single normal distributions turned out to be insufficient at higher statistics [69]. STEREO phase-I data were reanalyzed using the new method. Because of the smaller statistics, in particular in the background sample, the standard deviations $\sigma_{l,i}$ of the IBD PSD distributions in phase-I were constrained for each cell to $F_l \sigma_{T,i} \pm \sqrt{6} \sigma_{\sigma_{T,i}}$ implemented as pull term. Here, $\sigma_{T,i}$ is the standard deviation of the IBD PSD distribution in the TG, $\sigma_{\sigma_{T,i}}$ its statistical uncertainty and $F_l$ a factor that accounts for the different PSD resolution of cell 4 ($F_4 = 1.25$) compared to the other cells ($F_l = 0.95$, $l \neq 4$) in phase-I.

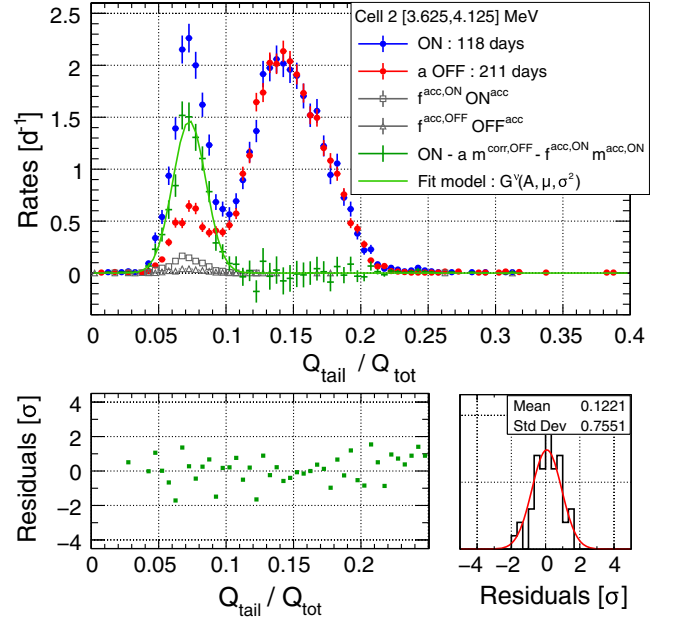

FIG. 27. Top: illustration of the extraction of the IBD rate for one energy bin of cell 2. Shown are the PSD distributions of prompt candidates and accidentals measured with reactor-on and reactor-off, where the latter is scaled by the parameter $a_{l,i}$ determined in the fit. The accidental distributions are also scaled by their respective factors $f^{acc,\xi}$, $\xi \in \{ON, OFF\}$. The green points show the difference between the $ON_{l,i}$ distribution on the one hand, and the sum of the models for the ON accidentals $f^{acc,ON} m_{l,i}^{acc,ON}$ and the OFF distribution $a_{l,i} m_{l,i}^{corr,OFF}$ on the other hand. It is fitted by a normal distribution with free normalization $G^\nu(A_{l,i}, \mu_{l,i}, \sigma_{l,i}^2)$ to determine the IBD rate. For the full fit model, see Eqs. (11)–(14) and the text. Note that the uncertainty bars according to the normal distribution are shown for reference only, since a likelihood fit with a Poisson distribution of the counts is used. Bottom: residuals of the fit model $G^\nu(A_{l,i}, \mu_{l,i}, \sigma_{l,i}^2)$ relative to the uncertainties of each data point. Individual residuals are shown in the left panel, while their distribution is shown in the right panel.

It is known that in the low statistics regime the estimator of the rate of IBDs can be biased [70]. The biases of each energy bin and cell are studied from simulations reproducing the fitting procedure and including all relevant experimental conditions (IBD rates, signal-over-background ratios, background shape, binning size etc.). For one pseudoexperiment, the bias is defined as the difference between the number of fitted antineutrinos and the number of generated antineutrinos, relative to the number of generated antineutrinos. Examining over 5000 pseudoexperiments, this quantity is distributed following a normal law from which the mean position is extracted and reported in Fig. 28. Below 6.5 MeV, the magnitude lies below 1%, while it reaches up to 2% in the last energy bin, where the ratio between signal and background events becomes smaller. For phase-I, with its lower statistics, the biases were typically below 2% but reached up to 6% in the last energy bin. The bias is corrected during the extraction of IBD candidates. In the case of the total spectrum, the PSD distributions of the six cells are summed before the fit leading to higher statistics and therefore a smaller bias, visible in Fig. 28 as black triangles.

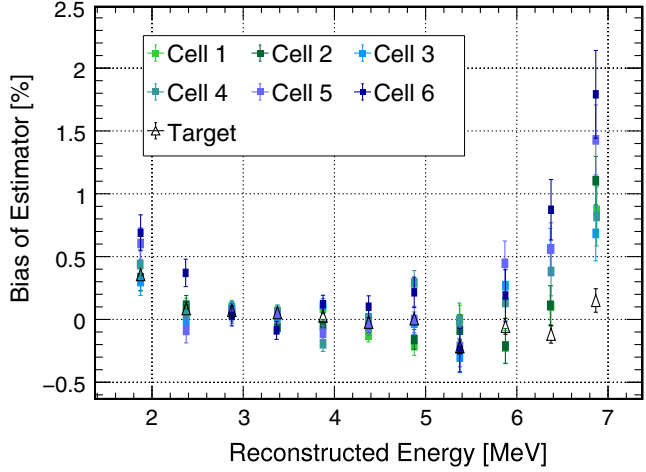

FIG. 28. Bias of the likelihood estimator of the IBD rates. Values are shown per cell and per energy bin, as well as for the total spectrum ("Target"). The estimates of the IBD rates are corrected from this bias.

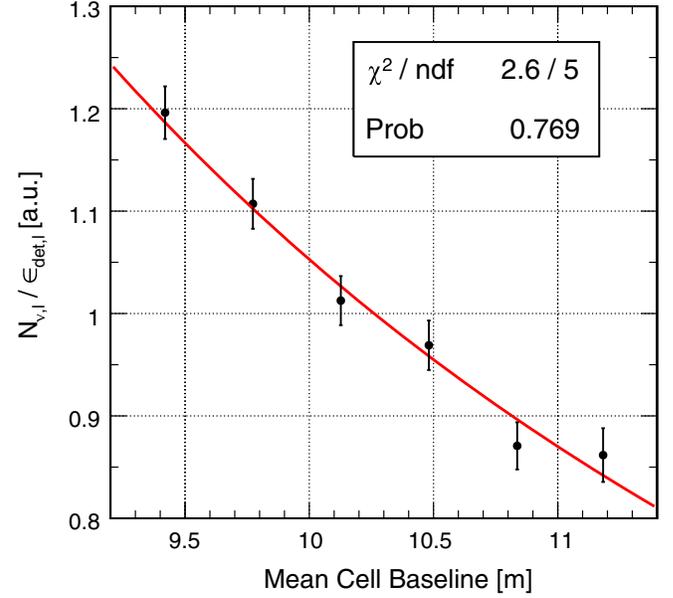

FIG. 29. Comparison of the measured rate of IBD events $N_{\nu,l}$ in each cell $l$ and integrated over all energy bins $i$ (black) with a fit of these data points using a $1/L^2$ model including a free normalization (red). For each cell, the mean cell baseline $L_l$ is used. It is calculated by taking into account the solid angle of incident antineutrinos. The data points have been corrected with the absolute detection efficiency $\epsilon_{\mathrm{det},l}$ for each cell.

## XI. OSCILLATION ANALYSIS

The oscillation analysis is performed in 11 equidistant energy bins of 500 keV width between 1.625 MeV and 7.125 MeV. For phase-I, the highest energy bin is excluded due to small statistics. The rate of measured IBD events for each of the six cells, extracted as described in Secs. VII and X, is compared with their respective expectations allowing for sterile neutrino oscillations. The main effect on the rate of IBDs in each cell is given by the dependence of the antineutrino flux on the distance $L$ from the reactor core. Figure 29 shows a very good agreement with a quadratic behavior when plotting the measured numbers of IBD events from all energy bins of cell $l$ over the mean cell baseline $L_l$.

To probe the signal of sterile neutrino oscillations, a relative comparison between cell spectra is performed via the following $\Delta\chi^2$ formalism:

$$
\chi^2 = \sum_{l=1}^{N_{\mathrm{cells}}} \sum_{i=1}^{N_{\mathrm{Ebins}}} \left( \frac{A_{l,i} - \phi_i M_{l,i}}{\sigma_{l,i}} \right)^2
$$
$$
+ \sum_{l=1}^{N_{\mathrm{cells}}} \left( \frac{\alpha_l^{\mathrm{EscaleU}}}{\sigma_l^{\mathrm{EscaleU}}} \right)^2 + \left( \frac{\alpha^{\mathrm{EscaleC}}}{\sigma^{\mathrm{EscaleC}}} \right)^2 + \sum_{l=1}^{N_{\mathrm{cells}}} \left( \frac{\alpha_l^{\mathrm{NormU}}}{\sigma_l^{\mathrm{NormU}}} \right)^2
$$
$$
\tag{15}
$$

with $l$ and $i$ being indices running over all cells and energy bins, respectively, $A_{l,i}$ denoting the measured IBD rates [cf. Eq. (11)] and $M_{l,i}$ being the corresponding expected IBD rates. The $M_{l,i}$ depend on the oscillation parameters and the set of nuisance parameters $\vec{\alpha}$.

Since we do not want to rely on absolute rate predictions in our analysis, the $\phi_i$ are introduced as free normalization parameters for each energy bin $i$. They effectively adjust the expected IBD rates $M_{l,i}$ across all cells $l$ to match the measured IBD rates $A_{l,i}$ on average. Since the $\phi_i$ absorb all absolute rate information per energy bin $i$, the analysis becomes independent of the spectrum prediction. Simultaneously with the $\phi_i$, the $M_{l,i}$ are optimized in terms of oscillation parameters [$\sin^2(2\theta_{ee})$, $\Delta m_{41}^2$] and nuisance parameters $\vec{\alpha}$ to match the remaining deviations from the $A_{l,i}$ in each cell $l$:

$$
M_{l,i} \equiv M_{l,i}(\sin^2(2\theta_{ee}), \Delta m_{41}^2, \vec{\alpha})
$$
$$
= \mathcal{M}_{l,i}(\sin^2(2\theta_{ee}), \Delta m_{41}^2) \cdot [1 + \alpha_l^{\mathrm{NormU}}
$$
$$
+ S_{l,i}^{\mathrm{Escale}} \cdot (\alpha_l^{\mathrm{EscaleU}} + \alpha^{\mathrm{EscaleC}})].
$$
$$
\tag{16}
$$

The dependence of $\mathcal{M}_{l,i}$ on the oscillation parameters is given by Eq. (1). This equation is applied individually to each MC event before it is registered in its corresponding bin $\mathcal{M}_{l,i}$. The parameters $\alpha^{\mathrm{EscaleC}}$ and $\alpha_l^{\mathrm{EscaleU}}$ account for the cell-to-cell correlated and uncorrelated energy scale uncertainties, respectively, while the additional nuisance parameters $\alpha_l^{\mathrm{NormU}}$ account for the cell-to-cell uncorrelated normalization uncertainties (cf. Table II). All nuisance parameters are constrained in Eq. (15) by their corresponding uncertainties via pull terms. Cell-to-cell correlated normalization uncertainties do not need to be included since the free normalization parameters $\phi_i$ compensate any cell-to-cell correlated normalization effect. $\alpha^{\mathrm{EscaleC}}$ and $\alpha_l^{\mathrm{EscaleU}}$ are considered in Eq. (16) via the sensitivity factors $S_{l,i}^{\mathrm{Escale}}$. They describe the sensitivity of the rate of IBD events in bin $i$ of cell $l$ to a distortion of the energy scale, i.e., the fraction of events received from or delivered to the neighboring energy bins of the cell, if its energy scale is changed by $(\alpha_l^{\mathrm{EscaleU}} + \alpha^{\mathrm{EscaleC}})$. The $S_{l,i}^{\mathrm{Escale}}$ are determined from a simulated energy spectrum of IBD prompt events scaled by the free normalization parameters $\phi_i$.

The statistical uncertainty

$$\sigma_{l,i} = \sigma_{l,i}(\phi_i M_{l,i}) \qquad (17)$$

in Eq. (15) is given as a function of the expected IBD rates $\phi_i M_{l,i}$. This is due to the fact that the scan of the parameter space of neutrino oscillations is a particular case in which the model deviates significantly from the experimental values when testing very large mixing angles. For this reason, $\sigma_{l,i}$ cannot be approximated by the statistical uncertainty of the data anymore and must evolve with the rate predicted by the model. The same simulation as the one developed for the study of likelihood biases (Sec. X) is used for this purpose. It estimates the statistical uncertainty of the IBD signal as a function of the expected rate, under real conditions of background rate and spectral shape.

This subtlety explains why the nuisance parameter $\alpha^{\mathrm{EscaleC}}$, describing cell-to-cell correlated shifts in the energy scale, is kept in the expression of the $\chi^2$ [cf. Eq. (15) and (16)] although it is, in principle, insensitive to correlated uncertainties. Here, the oscillation parameters induce different spectral distortions in each cell. Hence, $\alpha^{\mathrm{EscaleC}}$, albeit common to all cells, propagates differently in the expected rates $M_{l,i}$ and associated statistical uncertainties $\sigma_{l,i}$.

The no-oscillation scenario was tested by generating $10^4$ pseudoexperiments under the assumption of no sterile neutrino oscillations. For each pseudoexperiment, pseudodata are generated as fluctuations around the expected nonoscillated values within their uncertainty. Then, the value of

$$\Delta\chi^2(\sin^2(2\hat{\theta}_{ee}), \Delta\hat{m}_{41}^2)$$
$$= \chi^2(\sin^2(2\hat{\theta}_{ee}), \Delta\hat{m}_{41}^2, \hat{\vec{\alpha}}) - \chi^2(\sin^2(2\theta_{ee}), \Delta m_{41}^2, \vec{\alpha}) \qquad (18)$$

is calculated, where all parameters are allowed to vary with exception of $\sin^2(2\hat{\theta}_{ee})$ and $\Delta\hat{m}_{41}^2$. They are the parameters of the model to be tested, e.g., they are equal to zero in the case of the no-oscillation hypothesis.

As explained in Sec. III, the performance of the STEREO detector has changed between phase-I and phase-II. Therefore, we treat phase-I and phase-II as two separate experiments. In Eq. (18), we use the same $\chi^2$ formula [cf. Eq. (15)] for each one separately, but minimizing the sum of the two at the same time:

$$\chi_{\mathrm{PI+PII}}^2 = \chi_{\mathrm{PI}}^2(\sin^2(2\theta_{ee}), \Delta m_{41}^2, \vec{\alpha}_{\mathrm{PI}}, \phi_i, \Phi_{\mathrm{PI}})$$
$$+ \chi_{\mathrm{PII}}^2(\sin^2(2\theta_{ee}), \Delta m_{41}^2, \vec{\alpha}_{\mathrm{PII}}, \phi_i). \qquad (19)$$

The changed response of the detector between the phases is taken into account in the simulation. As the systematic uncertainties evolved between the two phases, the nuisance parameters $\vec{\alpha}$ are different. However, we keep the same normalization parameters $\phi_i$ as we expect the adjustments of the spectrum to be the same for the two phases. To be able to do so, we have to add a normalization term $\Phi_{\mathrm{PI}}$ which is common to all cells. We implement it for phase-I, where the parameters $\phi_i$ are replaced by the expression $\phi_i\Phi_{\mathrm{PI}}$ in Eqs. (15) and (17), respectively. The normalization parameter $\Phi_{\mathrm{PI}}$ is adjusted freely during the minimization and takes into account the different antineutrino rates for the two datasets caused by different average reactor powers. The parameters of interest $\sin^2(2\theta_{ee})$ and $\Delta m_{41}^2$ are kept identical for the two phases.

To test the no-oscillation hypothesis, we compare the $\Delta\chi^2$ of the data from phase-I + II with the distribution obtained from $10^4$ pseudoexperiments [71]. We get a p-value of 9%, i.e., the no-oscillation hypothesis cannot be rejected. Figure 30 compares the spectra measured in phase-II to the no-oscillation model fitted to phase-I + II data. We report the best-fit values of all nuisance parameters in Fig. 31. They are all contained within one standard deviation of their central values. For the free normalization parameter between the two phases, we get a value of $\Phi_{\mathrm{PI}} = 1.14 \pm 0.02$. As expected, it is consistent with the ratio of the average reactor powers of each phase, found to be $1.12 \pm 0.02$ (cf. Fig. 3).

While computing the $\Delta\chi^2$ values of the pseudoexperiments, we find that the fitting procedure tends to describe the generated pseudodata using values of $\sin^2(2\theta_{ee})$ and $\Delta m_{41}^2$, whose distribution roughly coincides with the shape of a sensitivity contour (cf. Fig. 32). Most of the values are in a parameter space region already excluded by global analyses [19–21]. We explain this by the high flexibility of the fit function [cf. Eq. (1)] which allows to describe statistical fluctuations present in the data. By going to high values of $\sin^2(2\theta_{ee})$ and $\Delta m_{41}^2$, the fit model is able to match individual fluctuations. In this context, best-fit points are expected to be distributed in the shape of a sensitivity curve, as too large (small) values of $\sin^2(2\theta_{ee})$ would induce an amplitude in Eq. (1) that is larger (smaller) than statistical fluctuations. As expected, the best-fit point of the real dataset also falls into this distribution (red star in Fig. 32). Comparing the goodness of fit of our best-fit point and the no-oscillation hypothesis in Table V, we find no strong preference for the best-fit point.

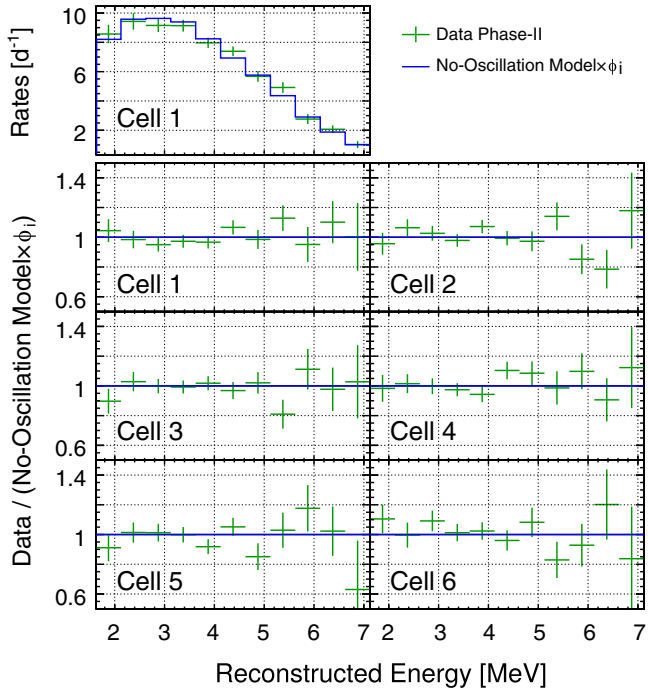

FIG. 30. Absolute (top) and relative (bottom) comparison between the measured rates of IBD events $A_{l,i}$ for phase-II in each cell $l$ (green) and the fitted no-oscillation model $M_{l,i}(0,0,\vec{\alpha})\phi_i$ (blue) after a simultaneous fit to phase-I + II data. All spectra are available in Ref. [72].

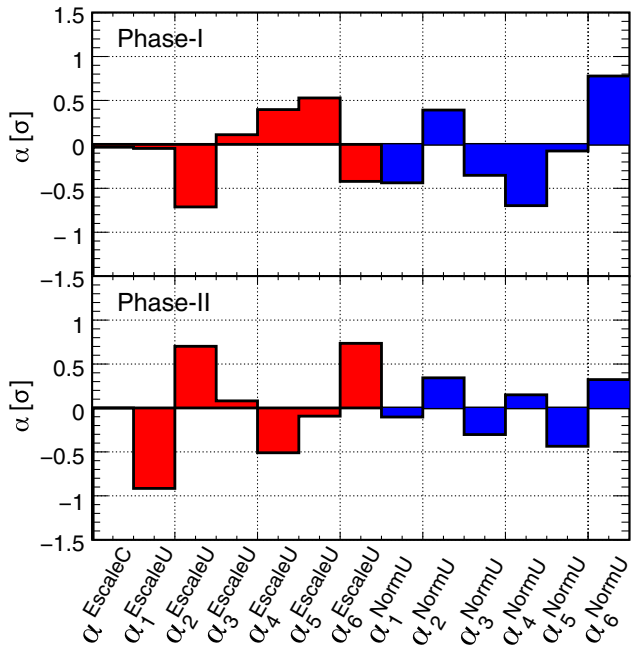

FIG. 31. Values of the nuisance parameters used in the oscillation fit [cf. Eq. (15)] for the best-fit to the no-oscillation model in phase-I + II. All parameters are contained within ±1 standard deviation with respect to their corresponding pull terms.

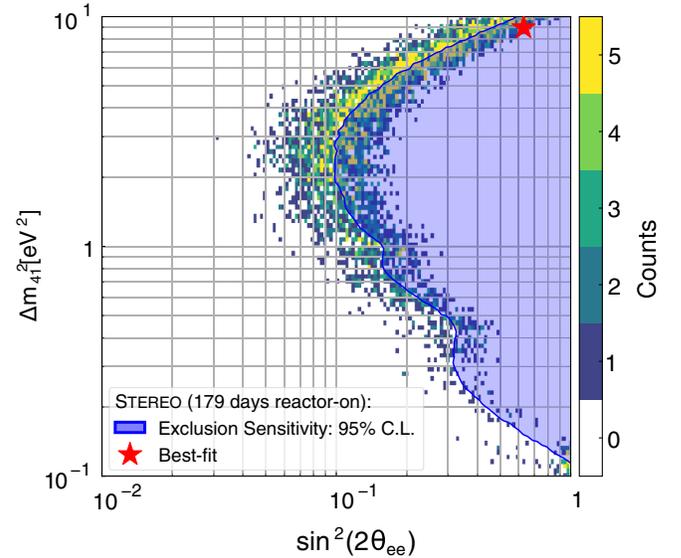

FIG. 32. Distribution of the best-fit values of 5000 pseudoexperiments generated under the no-oscillation hypothesis. The distribution has the shape of a sensitivity contour. Most best-fits are located in a region of high values of $\sin^2(2\theta_{ee})$ and $\Delta m_{41}^2$. The best-fit of the real dataset is $[\sin^2(2\theta_{ee}) = 0.63$, $\Delta m_{41}^2 = 8.95\ eV^2]$ (star) and falls into this region, as well.

When introducing a constraint on the normalization of the overall spectrum, the fit model no longer employs high values of $\sin^2(2\theta_{ee})$ and $\Delta m_{41}^2$ to describe the pseudodata. For this study, we introduced a pull term similar to the ones in Eq. (15), where the constrained parameter is the sum of the free normalization parameters $\phi_i$ across all cells [cf. Eq. (15)]. Instead of constraining this sum to the integral of the predicted IBD spectrum presented in Sec. IV, we constrain this sum, for each pseudoexperiment, to the sum of the free normalization parameters calculated for the pseudoexperiment under the assumption of no oscillations. In this study, we only impose a mild constraint of 20% relative uncertainty with respect to the reference. This constraint is chosen to be much looser than it would be possible from the uncertainty of the absolute rate. Note that we utilized the normalization constraint only for this study of best-fit point positions, but not in the determination of any oscillation result presented in this article.

TABLE V. Goodness of fit and hypothesis rejection for the no-oscillation hypothesis, the RAA best-fit point [13], and the STEREO best-fit point (cf. Fig. 32). All p-values are derived from comparisons of the experimental data with MC-distributions of $10^4$ pseudodatasets. See Eqs. (18) and (19) for the definitions of $\chi^2$ and $\Delta\chi^2$, respectively. The degrees of freedom (n.d.f.) are derived from the 6 cells and 11 (10) energy bins in phase-II (phase-I) on the one hand, and the 11 (1) bin-wise (phase-wide) free normalization parameters and 2 oscillation parameters on the other hand. Note that for the STEREO best-fit point, $\Delta\chi^2 = 0$ and p = 1 by definition [cf. Eq. (18)].

| | Goodness of fit | | Hypothesis test | |
|---|---|---|---|---|
| | $\chi^2$/n.d.f. | p-value | $\Delta\chi^2$ | p-value |
| No-oscillation | 137.4/114 | 0.07 | 9.0 | 0.09 |
| STEREO best-fit | 128.4/112 | 0.09 | 0 | 1 |
| RAA best-fit | 159.2/114 | 0.003 | 30.8 | $<10^{-3}$ |

To draw exclusion contours, we use the two-dimensional approach given in Eqs. (18) and (19). Since the conditions for applying Wilks' theorem [73] are not always met in sterile neutrino searches (cf. e.g., [74] for a discussion), we cannot assume standard $\chi^2$ distributions in the calculation of our exclusion contours. Instead, we obtain correct distributions by generating $10^4$ pseudodatasets in each parameter space point. Note that our approach yields much weaker exclusion contours as a simple application of Wilks' theorem [75]. The set of all rejected points is depicted in Fig. 33. In particular, we reject the best-fit point of the RAA at more than 99.9% C.L.

The oscillatory structure of the exclusion contour (red) around the expected mean sensitivity (blue) in Fig. 33 can be interpreted as statistical fluctuation. These structures are also found in similar measurements by other experiments [14–17]. By replacing the real dataset with several pseudodatasets, we are able to generate pseudoexclusion contours which show a similar oscillatory structure around the expected mean sensitivity, but whose extrema are located at different values of $\Delta m_{41}^2$. In the same way, we find the pseudoexclusion contours shifted along $\sin^2(2\theta_{ee})$ to either side of the sensitivity contour. This behavior is expected for the two-dimensional approach, when the no-oscillation hypothesis cannot be excluded by the data, where shifts toward lower $\sin^2(2\theta_{ee})$ are related to low p-values of the no-oscillation hypothesis [76]. In Fig. 34, we compare our sensitivity and exclusion contours presented in Fig. 33 with contours obtained with two alternative statistical methods. First, we compare them with the contours from a one-dimensional method, where rejected intervals of $\sin^2(2\theta_{ee})$ are obtained for a series of several fixed $\Delta m_{41}^2$ values. By plotting the rejected intervals for each $\Delta m_{41}^2$ value along the ordinate of Fig. 34, we achieve the depicted sensitivity and exclusion contours (raster-scan method). In addition, we compute the contours in a two-dimensional frequentist approach by normalizing the confidence level of the oscillation-hypothesis to the confidence level of the null-hypothesis, i.e., no-oscillation-hypothesis, (CLs method) [76,77]. As shown in Fig. 34, the sensitivities of all three methods agree across the relevant range of $\Delta m_{41}^2$, with the two-dimensional approach yielding a slightly weaker sensitivity. The exclusion contours also show, in general, good agreement with respect to their structure and position. However, for our dataset, the two-dimensional approach exhibits a stronger exclusion contour. As discussed above, this is expected, if the no-oscillation hypothesis cannot be excluded, but has a low p-value [76]. We note that the different exclusion and sensitivity contours are valid results within their respective method. For each method, results of different experiments can be compared with each other. Global analyses so far used two-dimensional results, which are particularly suited to derive allowed confidence regions. We therefore consider our two-dimensional result (cf. Fig. 33) as our main result, noting that also the CLs method allows for combined analyses.

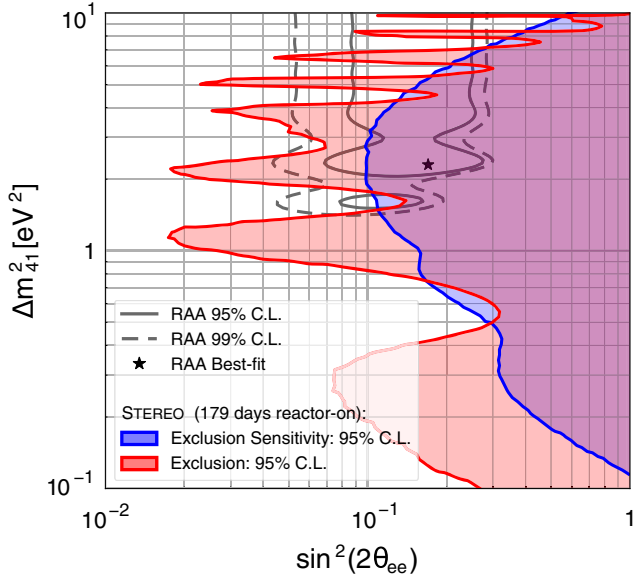

FIG. 33. Exclusion contour (red) and exclusion sensitivity contour (blue) at 95% C.L. of phase-I + II. Overlaid are the allowed regions of the RAA (grey) and its best-fit point (star) [13]. The contours and their underlying data are available in Ref. [72].

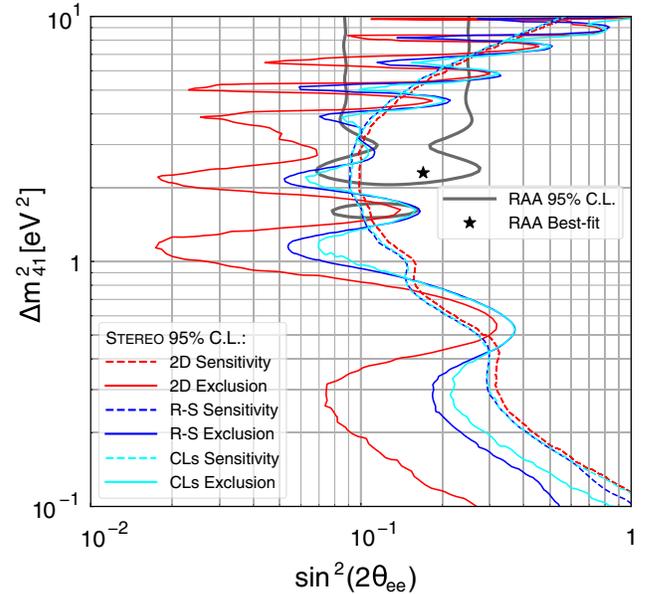

FIG. 34. Comparison of the exclusion contours (solid) and exclusion sensitivity contours (dashed) at 95% C.L. of phase-I + II, for the two-dimensional method (red, same as Fig. 33), the raster-scan method (dark blue), and the CLs method (light blue). Overlaid are the allowed regions of the RAA (grey) and its best-fit point (star) [13]. The contours and their underlying data are available in Ref. [72].

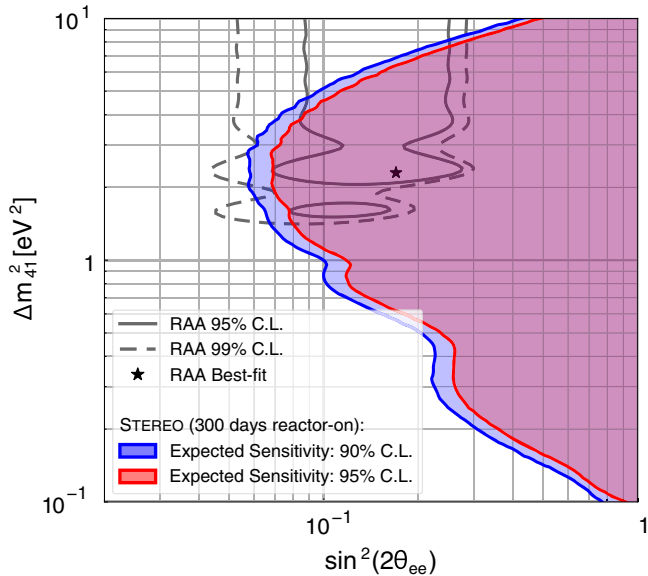

FIG. 35. Exclusion sensitivity contour at 90% C.L. (blue) and 95% C.L. (red) expected at the full dataset of the STEREO experiment. The expected number of days with reactor-on (off) is 300 (500) days. Overlaid are the allowed regions of the RAA (grey) and its best-fit point (star) [13].

To predict the sensitivity expected at the end of the STEREO experiment, we use the statistical uncertainties, calculated as described before, for the number of days expected at the end of the experiment. The expected number of days with reactor-on (off), normalized to the nominal reactor power, is 300 (500) days. The sensitivity contours are then computed by the aforementioned raster-scan method, which is the least computationally expensive. The resulting sensitivity is depicted in Fig. 35. It is calculated assuming the systematic uncertainties derived from the data acquired until now. It may improve in the future if systematic uncertainties further reduce. We expect the statistical uncertainties to be comparable to the systematic uncertainties in the final dataset of the STEREO experiment.

All results of the approach given by Eq. (15) have been confirmed by an independent $\Delta\chi^2$ method, where all uncertainties are modelled by a covariance matrix instead of nuisance parameters [78].

The exclusion and mean sensitivity contours of the three methods along with maps of $\Delta\chi^2$ and critical $\Delta\chi^2$ values at different confidence levels are available as Supplementary Material in Ref. [72]. When combining with other experimental data, special care should be exercised. Depending on the method, the assumption of a standard $\chi^2$ law (Wilks' theorem [73]) instead of the provided critical $\Delta\chi^2$ values derived from nonstandard $\chi^2$ distributions leads to modified contours [75]. In addition, for the interpretation of derived allowed confidence regions, one should always consider additional available information, in particular rate information. This is especially advisable considering the expected distribution of best-fit points in a prediction-free shape-only analysis, as shown in Fig. 32 for the STEREO experiment. More technical details are provided in Ref. [72].

## XII. CONCLUSION

After the initial results of phase-I of the STEREO experiment [18] based on 66 days of reactor-on data, the dataset was extended to include in total 179 days of reactor-on and 235 days of reactor-off data in phase-I + II. During the long reactor shutdown between the two phases, the optical separation walls of the detector were largely improved and the internal calibration system of the detector enhanced. In addition, several aspects of the data analysis and evaluation of systematic uncertainties were improved. The optical model was updated for phase-II and the optical properties of various detector components were refined. Here, special emphasis was given to the modeling of separation walls and the light cross-talk between the cells of the TG volume. The accuracy of the simulation of the reactor flux and energy spectrum was improved by utilizing more precise calculations of corrections to the Huber model. Special care was given to the dominant contribution from antineutrinos originating from $\beta$-decays of $^{28}$Al and $^{56}$Mn across the entire reactor heavy water tank. Due to an improved optical model in the detector simulation as well as the exploitation of the energy spectrum of $^{12}$B events, the energy scale of the experiment underwent a reduction of uncertainty across the entire energy range and TG volume. The selection uncertainties of IBD events were studied in great detail and corrections to the MC model were derived. In particular, a three-dimensional correction map for the neutron selection efficiency in the entire TG was determined, correcting for an imperfect description of the neutron mobility in the simulations. In addition, a more precise simulation of the $\gamma$-cascades after neutron captures on Gd exploiting the new FIFRELIN simulation package was developed. Systematic uncertainties in the background description were reduced by fitting in-situ measured PSD distributions of background acquired in reactor-off and reactor-on phases. Various systematic effects in this method were investigated and are accounted for accordingly. The statistical oscillation analysis of the resulting IBD spectra was performed by a prediction-independent procedure with free normalization factors which is insusceptible to cell-to-cell correlated systematics. We exclude the best-fit point of the RAA at more than 99.9% C.L. by combining our phase-I and phase-II datasets on a statistical basis using the aforementioned method in a full two-dimensional analysis. For our expected final dataset, we estimate our final 95% C.L. exclusion sensitivity to fully cover the relevant $\Delta m^2$ range of the 95% allowed region of the RAA.

To allow inclusion of this work into global oscillation analyses and further works, we provide various results of our analysis in digitized form. These supplements along with technical instructions can be found in Ref. [72].

## ACKNOWLEDGMENTS

This work is funded by the French National Research Agency (ANR) within the Project No. ANR-13-BS05-0007 and the "Investments for the future" programmes P2IO LabEx (ANR-10-LABX-0038) and ENIGMASS LabEx (ANR-11-LABX-0012). Authors are grateful for the technical and administrative support of the ILL for the installation and operation of the STEREO detector. We thank G. Mention and F. Le Diberder for valuable input on the statistical analyses and A. Onillon for running the TRIPOLI simulation at the origin of Fig. 5 and associated results. We further acknowledge the support of the CEA, the CNRS/IN2P3 and the Max Planck Society.